\documentclass[]{article}
\usepackage[margin=1in]{geometry}

\usepackage[utf8]{inputenc}
\usepackage[pdftex, pdftitle={Article}, pdfauthor={Author}]{hyperref} % For hyperlinks in the PDF
\usepackage[parfill]{parskip}
\usepackage{float} 
\usepackage{array}
\usepackage{multirow}
\usepackage[left]{lineno}
\usepackage{siunitx}
\usepackage{afterpage}
\usepackage{multicol}
\usepackage{multirow}
\newcolumntype{C}{>{\centering\arraybackslash}p{1 cm}}
\newcolumntype{V}{>{\centering\arraybackslash}p{2 cm}}
\newcolumntype{B}{>{\centering\arraybackslash}p{2.8 cm}}
\usepackage{chemformula}
\usepackage{graphicx}
\usepackage{scalerel,amsmath,amssymb,amsthm}
\usepackage[
backend=biber,
style=nature,
citestyle=numeric-comp,
sorting=none
]{biblatex}
\usepackage{bm}
\newcommand{\um}{$\mu$m }

\usepackage[utf8]{inputenc}
\usepackage{setspace} 
\usepackage{authblk}

\title{\textbf{
Kinetic Inductance of Few-Layer \ch{NbSe2} in the Two-Dimensional Limit}}
\date{}
\author[1,2]{Sameia Zaman}
\author[1*]{Joel~\^I-j.~Wang}
\author[3]{Thomas Werkmeister}
\author[1]{Miuko Tanaka}
\author[4]{Thao Dinh}
\author[1]{Max Hays}
\author[1]{Daniel Rodan-Legrain}
\author[1]{Aranya Goswami}
\author[1]{R\'eouven Assouly}
\author[4]{\\Ahmet Kemal Demir}
\author[5]{David K. Kim}
\author[5]{Bethany M. Niedzielski}
\author[1,5]{Kyle Serniak}
\author[5]{\\Mollie E. Schwartz}
\author[6]{Kenji Watanabe}
\author[7]{Takashi Taniguchi}
\author[3,8]{Philip Kim}
\author[4]{\\Riccardo Comin}
\author[1]{Jeffrey A. Grover}
\author[1,2]{Terry P. Orlando}
\author[4]{Pablo Jarillo-Herrero}
\author[1,2,4*]{William D. Oliver}

\affil[1]{Research Laboratory of Electronics, Massachusetts Institute of Technology, Cambridge, MA 02139, USA}
\affil[2]{Department of Electrical Engineering and Computer Science, Massachusetts Institute of Technology, Cambridge, MA 02139, USA}
\affil[3]{John A. Paulson School of Engineering and Applied Sciences, Harvard University, Cambridge, MA, USA}
\affil[4]{Department of Physics, Massachusetts Institute of Technology, Cambridge, MA 02139, USA}
\affil[5]{Lincoln Laboratory, Massachusetts Institute of Technology, Lexington, MA 02421, USA}
\affil[6]{Research Center for Electronic and Optical Materials, National Institute for Materials Science, 1-1 Namiki, Tsukuba 305-0044, Japan}
\affil[7]{Research Center for Materials Nanoarchitectonics, National Institute for Materials Science,  1-1 Namiki, Tsukuba 305-0044, Japan}
\affil[8]{Department of Physics, Harvard University, Cambridge, MA, USA}

\begin{document}

\maketitle
% \normalsize{$^\dagger$These authors contributed equally to this work.}\\
\normalsize{$^\ast$To whom correspondence should be addressed: \textcolor{blue}{joelwang@mit.edu,} and \textcolor{blue}{william.oliver@mit.edu}}
\pagebreak
\section*{Abstract}
\textbf{Van der Waals (vdW) superconductors remain superconducting down to the monolayer limit, enabling the exploration of emergent physical phenomena and functionality driven by reduced dimensionality. Here, we report the characterization of the kinetic inductance of atomically thin \ch{NbSe2}, a two-dimensional van der Waals superconductor, using superconducting coplanar waveguides and microwave measurement techniques familiar to circuit quantum electrodynamics (cQED). The kinetic inductance scales inversely with the number of \ch{NbSe2} layers, reaching 1.2 nH/$\bm{\Box}$ in the monolayer limit. Furthermore, the measured kinetic inductance exhibits a thickness-dependent crossover from clean- to dirty-limit behavior, with enhanced dirty-limit contributions emerging in the ultra-thin regime. These effects are likely driven by increased surface scattering, multi-band superconductivity, and geometric confinement. 
Additionally, the self-Kerr nonlinearity of the \ch{NbSe2} films ranges from $\bm{K/2\pi}$ = –0.008 to –14.7 Hz/photon, indicating its strong potential in applications requiring compact, nearly linear, high-inductance superconducting quantum devices and detectors. The fabrication and characterization techniques demonstrated here are extensible to the investigation of other two-dimensional superconductors.}

% These findings establish \ch{NbSe2} as a material that combines superconductivity, large kinetic inductance, and low disorder, making it suitable for building lumped-element superconducting quantum devices and detectors. 

\section*{Introduction}
Kinetic inductance ($L_\mathrm{k}$) in a superconductor arises from the inertia of Cooper pair transport in response to a varying electromagnetic field. Unlike geometric inductance, which varies with the geometry of the electronic element, kinetic inductance depends on intrinsic properties of the material, e.g., superfluid density, carrier effective mass, and film thickness. For example, in a clean superconductor, the kinetic inductance per square is $L_\mathrm{k,sq}  = m_\mathrm{eff}/ e^2n_\mathrm{0}d$, where $m_\mathrm{eff}$ is the effective mass of the charge carriers, $n_\mathrm{0}$ is the superfluid density (number of Cooper pairs per unit volume) per unit thickness, and \textit{d} is the thickness of the sample. This relation indicates that $L_\mathrm{k,sq}$ increases with decreasing film thickness or suppressed superfluid density. Such an enhancement is useful in quantum circuits, where superconductors with high-$L_\mathrm{k}$ are used to realize superinductors --- inductive elements with a reactance that approaches and even exceeds the quantum unit $R_\mathrm{Q}=h/(2e)^2 = \SI{6.4}{k\Omega}$ for Cooper pairs, where $h$ is the Planck constant, and $e$ is the elementary charge. Furthermore, the kinetic inductance in a superconductor is directly related to the superfluid stiffness; probing it enables insight into several fundamental properties such as the superconducting gap structure and pairing symmetry in unconventional superconductors \cite{phan2022detecting, bottcher2024circuit, kreidel2024measuring, tanaka2025superfluid, banerjee2025superfluid, jin2025exploring}.

Superinductors are used in a variety of applications, including in superconducting qubit circuits such as fluxonium and 0-$\pi$ qubits, which require large (and ideally linear) shunt inductance~\cite{manucharyan2009fluxonium, gyenis2021experimental}. They have also been used to realize coherent quantum phase slip circuits~\cite{astafiev2012coherent,shaikhaidarov2022quantized}, long-range coupling for spin qubits~\cite{mi2018coherent}, and tailored Kerr non-linearities in light-matter interactions~\cite{ye2025near,puertas2019tunable}. Furthermore, these materials can be used for wide-band parametric amplifiers and high-impedance readout resonators~\cite{bottcher2022parametric, clerk2020hybrid, landig2018coherent, mantegazzini2023high, splitthoff2024gate}. For many of these applications, low-$L_\mathrm{k}$ aluminum is used, and thus, large inductance is realized primarily by an array of Al-\ch{AlO$_x$}-Al Josephson junctions \cite{macklin2015near, nguyen2019high, crescini2023evidence}. However, these junctions exhibit strong nonlinearity and are therefore limited in the current they can support while remaining in the linear inductance regime. This motivates the search for alternative materials that can provide large inductance with minimal nonlinearity. Recent efforts toward this goal have focused on implementing superinductors using amorphous superconducting thin films, such as TiN~\cite{shearrow2018atomic,amin2022loss}, NbN~\cite{frasca2023nbn,niepce2019high}, NbTiN~\cite{samkharadze2016high, bretz2022high}, \ch{InO$_x$}~\cite{astafiev2012coherent,dupre2017tunable}, NbSi~\cite{calvo2014niobium}, WSi \cite{kirsh2021linear}, \ch{AlO$_x$} \cite{zhang2019microresonators}, and granular Al \cite{grunhaupt2018loss, glezer2020granular}. The inherent disorder of these materials suppresses the density of Cooper-pairs, leading to high-$L_\mathrm{k}$ in the superconducting regime \cite{Charpentier2025}. Typically, these amorphous superconductors operate in the dirty limit, where the mean free path of charge carriers is shorter than the superconducting coherence length \cite{tinkham2004introduction}. This relationship implies that, in the BCS framework, the size of a Cooper pair is greater than the average travel distance between consecutive scattering sites. In contrast, the coherence length of a clean superconductor is shorter than its mean free path, and Cooper pair transport can be regarded as ballistic when the length of the device is shorter than the mean free path.

In superconducting quantum devices, disorder and the resulting scattering of charge carriers are expected to affect quasiparticle dynamics and, consequently, device coherence. While the link between disorder, quasiparticle generation, and the associated noise remains an area of active investigation \cite{muller2019towards, grunhaupt2018loss, arute2019quantum, kurilovich2025correlated}, microwave characterization of superconductors with varied disorder levels may offer a way to probe their dynamic responses and help explore the role of scattering in decoherence~\cite{kurilovich2025correlated}. %In this work, we find that the kinetic inductance of encapsulated \ch{NbSe2} scales inversely with its thickness, reaching as high as \SI{1.2}{\nano\henry}/$\Box$ approaching the monolayer limit, comparable to the highest values reported for high-impedance superconductors.

% In superconducting quantum devices based on dirty superconductors, scattering between Cooper-pairs and disorder is associated with excessive quasiparticle generation and noise, potentially limiting device coherence times \cite{muller2019towards, grunhaupt2018loss, arute2019quantum, kurilovich2025correlated}. Studying clean superconductors using a cQED platform may offer a means to probe these decoherence mechanisms in the opposite regime, with the potential to reduce energy loss and suppress quasiparticle-induced decoherence \cite{kurilovich2025correlated}.
%\ch{NbSe2} presents an opportunity to study fundamental characteristics of kinetic inductance in the opposite regime.   

\section*{\ch{NbSe2}--a van der Waals Superconductor}
We investigate 2H-\ch{NbSe2} (further referred to as \ch{NbSe2} for brevity), a transition-metal dichalcogenide and a type-II superconductor. The lattice structure of \ch{NbSe2} is hexagonal, with the Nb atoms in trigonal prismatic coordination surrounded by Se atoms (Fig.~\ref{fig:method}a). The layers are stacked in an ABAB sequence along the c-axis and are bonded by van der Waals forces between adjacent layers. \ch{NbSe2} possesses out-of-plane mirror symmetry and broken in-plane inversion symmetry. This broken in-plane symmetry gives rise to an unconventional Ising spin-orbit coupling, which maintains superconductivity in the presence of a strong in-plane magnetic field exceeding \SI{30}{\tesla}, as observed in the monolayer limit \cite{xi2016ising, de2018tuning}. Both superconductivity and the charge-density wave transition exist from the bulk down to monolayer thickness. %with the transition characteristics influenced by the film’s dimensionality. 
However, the superconducting transition temperature ($T_\mathrm{c}$) decreases with reduced film thickness, attributable to a suppressed Cooper-pair density, the presence of a superconductor–vacuum interface, and modifications to the electronic band structure \cite{khestanova2018unusual, cao2015quality}. As the thickness is reduced, \ch{NbSe2} exhibits several unusual quantum phenomena—including thickness-dependent Higgs modes \cite{du2025unveiling}, dissipationless phase diagrams \cite{benyamini2019fragility}, unconventional finite momentum
pairing \cite{wan2023orbital,hamill2021two,cho2022nodal}, and the emergence of a quantum metallic phase under a small perpendicular magnetic field \cite{tsen2016nature}. %These features emerge distinctly in the disorder-free clean superconductors. 
Prior direct current (DC) transport studies indicate that the carrier mean free path in \ch{NbSe2} decreases with its thickness due to enhanced surface scattering \cite{khestanova2018unusual, benyamini2019fragility}. When encapsulated by hBN, even bilayer and monolayer \ch{NbSe2} exhibit mean free paths (approximately \SI{17}{\nm} and \SI{15}{\nm}, respectively) that are larger than their corresponding superconducting coherence length (\SI{8}{\nm}) \cite{khestanova2018unusual, de2018tuning, wan2023orbital}. These DC transport measurements are consistent with encapsulated \ch{NbSe2} exhibiting superconductivity in the clean limit \cite{prober1980upper, khestanova2018unusual, zhu2022transport, benyamini2019blockade}. 

% The ability to tune physical properties through dimensionality reduction and disorder control positions \ch{NbSe2} as an ideal platform for exploring superconductivity, confinement-induced emergent phenomena, and potential device applications in the clean limit \cite{saito2016highly}. 

While previous studies have primarily relied on DC transport to probe only the real part of the conductivity, in this work, we measure the kinetic inductance to access the complex conductivity of few-layer \ch{NbSe2} in the two-dimensional limit. This method allows us to determine whether the material lies in the clean limit, the dirty limit, or an intermediate regime. Moreover, it is sensitive to fundamental superconducting properties such as superfluid stiffness, surface impedance, and loss mechanisms \cite{tanaka2025superfluid, maji2024superconducting, wang2022hexagonal}, providing insight into the material suitability for superconducting quantum devices operating in the microwave regime.

\section*{Device Design}
\begin{figure}[H]
    \centering
    \includegraphics[width= \textwidth] {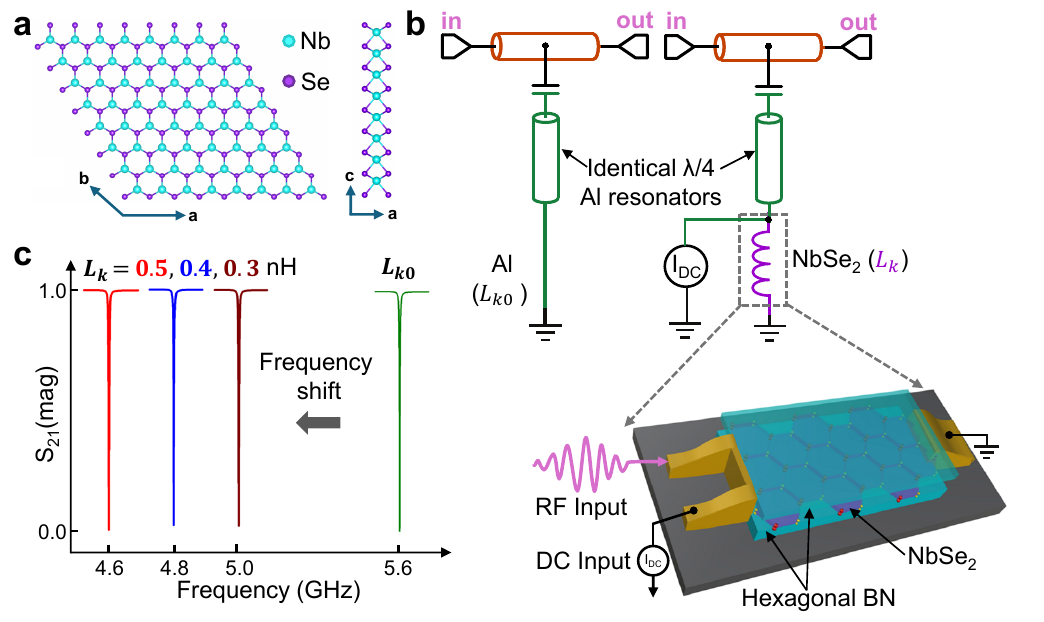}
    \caption{\textbf{Superconducting resonators and terminations to characterize the kinetic inductance of \ch{NbSe2}.} \textbf{a,} Crystal structure of monolayer 2H-\ch{NbSe2}, top-down and side views. \textbf{b,} Measurement circuit schematic. CPW resonators are capacitively coupled to a common transmission line and terminated either directly to ground via aluminum (control) or through a hBN-\ch{NbSe2}-hBN heterostructure (experiment). The resonant frequency $f_\mathrm{r}$ of the aluminum $\lambda$/4 resonator is determined by the effective inductance and capacitance of the CPW. Termination by \ch{NbSe2} introduces additional kinetic inductance, shifting the resonant frequency to a new value $f_\mathrm{r'}$. (Zoom-in of inductor) hBN-\ch{NbSe2}-hBN heterostructure contacted by Al electrodes with both RF and DC input signals applied across the sample and terminated at ground. \textbf{c,} Simulated resonator spectra illustrating the effect of increased inductance on the resonant frequency. $L_\mathrm{k0}$ denotes the inductance of the aluminum-terminated resonator, while $L_\mathrm{k}$ represents the additional inductance introduced by the \ch{NbSe2} sample.}
    \label{fig:method}
\end{figure}

We study the kinetic inductance of hBN-encapsulated \ch{NbSe2} thin films using quarter-wavelength $\lambda/4$ coplanar waveguide (CPW) resonators made of thin-film Al. The resonant frequency of a resonator is $f_\mathrm{r} = \frac{1}{2\pi {\sqrt{L_{\text{eff}}C_{\text{eff}}}}}$, where $L_{\text{eff}}$ and $C_{\text{eff}}$ are the effective inductance and capacitance of the resonator, respectively (Fig.~\ref{fig:method}b, left panel). The Al inductance is primarily geometric, and its kinetic inductance per square is much smaller than that of \ch{NbSe2}. When the $\lambda/4$-resonator is terminated to ground by an \ch{NbSe2} sample, the resonant frequency shifts to a lower frequency $f_\mathrm{r'} = \frac{1}{2\pi \sqrt{L'_{\text{eff}}C'_{\text{eff}}}}$ due to the additional kinetic inductance from the \ch{NbSe2} (Fig.~\ref{fig:method}b, right panel). We perform microwave simulations to determine the expected frequency shift of a $\lambda/4$-resonator due to the additional inductance $L_\mathrm{k}$ (Fig.~\ref{fig:method}c). The Al-only resonator has a frequency of \SI{5.6}{\giga \hertz}. When terminated by \ch{NbSe2}, the resonance frequency decreases by 200 MHz per \SI{0.1}{nH} of added termination inductance.

% which decreases by \SI{200}{\mega \hertz} for each additional \SI{0.1}{\nano \henry} of $L_\mathrm{k}$. %Since the superfluid density of \ch{NbSe2} decreases with the number of layers, we can use this technique to map the behavior of kinetic inductance across various layer thicknesses of \ch{NbSe2}. 

\begin{figure}[H]
    \centering
    \includegraphics[width= \textwidth]{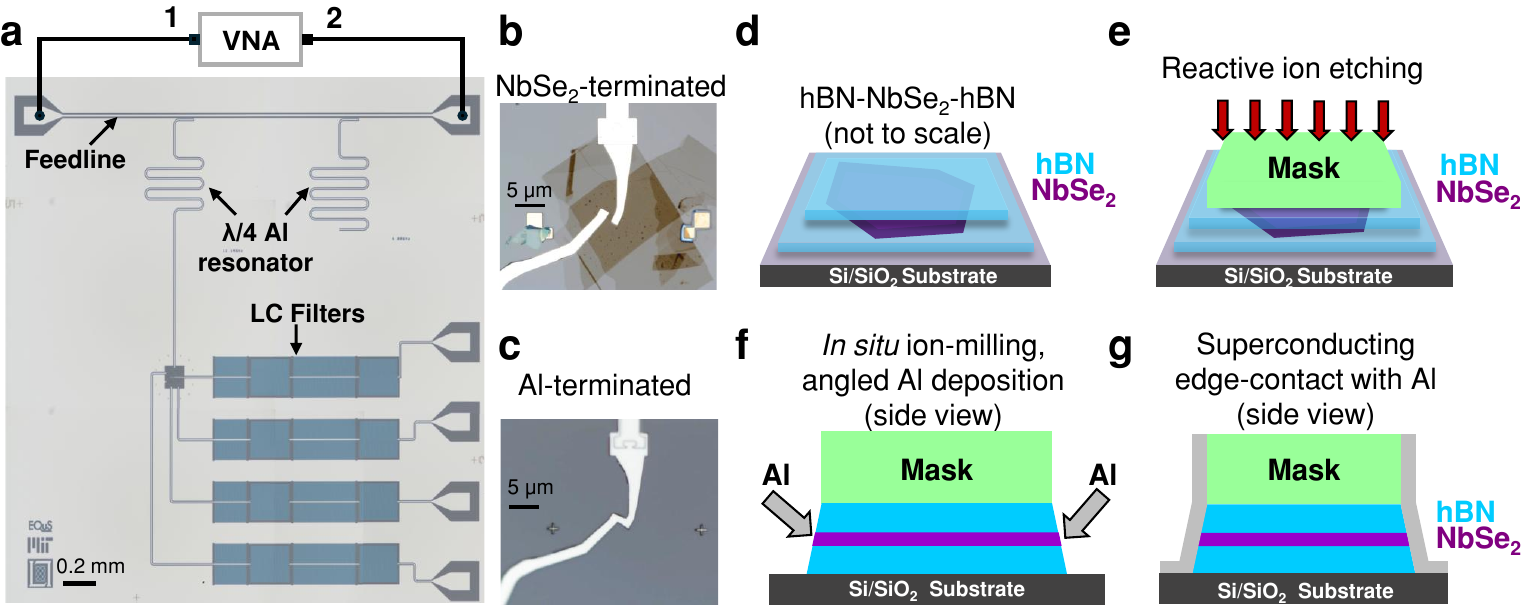}
    \caption{\textbf{Measurement configuration and device fabrication for the kinetic inductance measurement of \ch{NbSe2}.} \textbf{a,} Optical micrograph of a 5 $\times$ \SI{5}{mm^2} chip containing CPW resonators, a shared feedline, DC bias lines, on-chip LC filters, and a ground plane, all patterned from \SI{250}{nm}-thick aluminum on a high-resistivity silicon substrate. Only one (of four) DC bias lines is used in this experiment. \textbf{b - c,} Optical micrographs of the \ch{NbSe2}-terminated and aluminum-terminated $\lambda$/4 CPW resonators, respectively. \textbf{d - g,} Illustration of the edge-contact fabrication process. The \ch{NbSe2} (purple) flake is fully encapsulated by hBN (blue) and patterned via reactive-ion etching (RIE) to expose its edges. After in-situ argon ion milling, superconducting edge contacts are formed by angled aluminum evaporation with substrate rotation.}
    \label{fig:device}
\end{figure}

Figure~\ref{fig:device}a shows an optical micrograph of the superconducting circuit with $\lambda$/4 CPW resonators used for this work. The aluminum CPW resonators are fabricated from a \SI{250}{\nano\meter} Al film deposited on a high-resistivity silicon substrate and patterned using a combination of photolithography and wet etching (see Ref.~\cite{tanaka2025superfluid} and Supplementary Information for details). Two $\lambda/4$-resonators, an aluminum \textit{spectator} resonator and an \textit{experiment} resonator, are capacitively coupled to a common feedline. The spectator resonator is used as a witness to monitor the impact of the vdW materials fabrication process (without incorporating vdW materials), while the experiment resonator is terminated by a \ch{NbSe2} sample situated within a square hosting window of size $\SI{200}{\micro\meter}$ by $\SI{200}{\micro\meter}$. To extract the kinetic inductance of each \ch{NbSe2} device, we analyze two $\lambda/4$-resonators: the \ch{NbSe2}-terminated experiment resonator (Fig.~\ref{fig:device}b) and a control resonator of the same geometry that is terminated instead by a \SI{250}{nm}-thick aluminum film (Fig.~\ref{fig:device}c). The \ch{NbSe2} thin film is encapsulated in hBN to maintain a nominally clean and atomically flat interface while ensuring a relatively low-microwave-loss environment \cite{Dean2010, wang2022hexagonal}. This hBN-\ch{NbSe2}-hBN heterostructure is assembled through mechanical exfoliation and standard dry-transfer techniques inside an argon-filled glove box, ensuring minimal oxidation of the \ch{NbSe2} before full encapsulation. 

% This device configuration and fabrication process enable the investigation of the intrinsic kinetic inductance of pristine two-dimensional superconductors. 

The hBN-\ch{NbSe2}-hBN heterostructure is galvanically connected to the microwave resonator and the ground plane. To achieve a highly transparent, superconducting contact, we make a one-dimensional edge-contact to the \ch{NbSe2} using reactive ion etching of the heterostructure, followed by in-situ argon ion milling and aluminum deposition (Fig.~\ref{fig:device}d) \cite{sinko2021superconducting}. A low-loss superconducting contact is critical to maintaining a narrow resonance line (i.e., a high quality factor of the resonator), which translates to higher measurement resolution of the resonator frequency. Details of the fabrication process and the verification of the superconducting contact are provided in the Supplementary Information.

% The superconducting contact is critical to ensure the microwave current is terminated to ground through the \ch{NbSe2} sample with minimal loss, enabling frequency measurements with high resolution (limited by the quality factor of the resonator)
\section*{Microwave Characterization of \ch{NbSe2}}
The samples are characterized in a dilution refrigerator with a base temperature of approximately \SI{10}{mK} (see Fig.~\ref{fig:meas_chain} for wiring diagram). The microwave transmission coefficient $S_\mathrm{21}$ is measured as a function of frequency using a two-port vector network analyzer (VNA). Figure~\ref{fig:measurement}a shows a representative $|S_\mathrm{21}|$ measurement from a $\lambda/4$-resonator terminated by a 9-nm-thick (determined by AFM) \ch{NbSe2} film (Device ID D6). 
The resonant frequency $f_\mathrm{r,Al-NbSe_2}$ and quality factor are determined from a Lorentzian fit using the standard circular fit procedure \cite{Probst2015} (Fig.~\ref{fig:measurement}a inset and Supplementary Information). In the single-photon limit, the \ch{NbSe2}-terminated resonator exhibits a resonant frequency $f_\mathrm{r,Al-NbSe_2}$ = \SI{4.79}{\giga \hertz} and an internal quality factor $Q_\mathrm{i,Al-NbSe_2} = 2.8\times10^4$, while the Al-terminated control resonator has a resonant frequency $f_\mathrm{r,Al}$ = \SI{5.34}{\giga \hertz} and an internal quality factor $Q_\mathrm{i,Al} = 1.0\times10^5$. The lower internal quality factor observed in the \ch{NbSe_2}-terminated Al resonator is likely due to additional dissipation arising from imperfect electrical contacts between the Al and the \ch{NbSe_2}, as well as dielectric loss introduced by residual polymer contamination from the dry transfer process. Moreover, the frequency difference $f_\mathrm{r,Al} - f_\mathrm{r,Al-NbSe_2}$ = \SI{0.45}{\giga \hertz} highlights the impact of terminating the resonator with \ch{NbSe2} and enables us to extract a kinetic inductance $L_\mathrm{k,sq}$ = \SI{124}{\pico \henry} for this \ch{NbSe2} sample. 

The resonant frequency of the \ch{NbSe2}-terminated resonator decreases with increasing microwave probe power, e.g., with a noticeable downshift in $f_\mathrm{r,Al-NbSe_2}$ at the estimated sample power of \SI{-90}{\decibel m} (Fig.~\ref{fig:measurement}b, green arrow). This frequency downshift indicates a contribution from kinetic inductance. As the microwave power increases, enhanced Cooper-pair breaking leads to a reduction in the superfluid density ($n_\mathrm{0}$), resulting in a larger kinetic inductance $L_\mathrm{k}$ and, consequently, a lower resonant frequency. In contrast, the aluminum resonator exhibits no observable power dependence up to an estimated input power of \SI{-60}{dBm} at the sample (Fig.~\ref{fig:meas_Al_shunted}b), corresponding to the upper limit of our VNA output (VNA outputs a power of \SI{10}{dBm} at room temperature, which is reduced to \SI{-60}{dBm} at the sample after passing through a total attenuation of \SI{70}{dB}), confirming that its inductance is predominantly geometric. The nonlinear microwave power dependence exhibited by the \ch{NbSe2}-terminated resonator, observed over the power range from \SI{-100}{dBm} to \SI{-80}{dBm}, is well-described by a Kerr-type or Duffing-type nonlinearity \cite{nayfeh2008nonlinear} (see Supplementary Information). When the microwave power in the resonator exceeds a critical threshold, the resonant frequency bifurcates and enters a bistability regime (Fig.~\ref{fig:measurement}b, blue arrow)~\cite{lee2007nonlinear, vijay2009invited, andersen2020quantum}. From a linear fit of the resonant frequency shift ($\Delta f = f_\mathrm{r,Al-NbSe_2}(n_\mathrm{r}= 1) - f_\mathrm{r,Al-NbSe_2}(n_\mathrm{r})$) with the photon number of the resonator ($n_\mathrm{r}$), shown in Fig.~\ref{fig:measurement}c, we extract a self-Kerr coefficient $K/2\pi = \SI{-1.7}{Hz}$/photon at \SI{10}{mK}. Typically, conventional superconducting inductors of similar dimensions, such as granular aluminum, with a self-Kerr coefficient of \SI{4.5}{MHz}~\cite{winkel2020implementation}, and Josephson junction arrays, with a self-Kerr of \SI{8}{MHz}~\cite{weissl2015kerr}—exhibit significantly higher nonlinearity. In contrast, \ch{NbSe2} films show a lower self-Kerr coefficient, indicating low nonlinearity.

We also investigate the dependence of the kinetic inductance on the DC bias current. A DC bias current $I_\mathrm{DC}$ is introduced via a contact to the microwave line just above the \ch{NbSe2} termination (Fig.~\ref{fig:measurement}d), while a microwave tone is maintained in the low-power linear regime ($P_\mathrm{rf} = \SI{-120}{dBm}$). The resonant frequency ($f_\mathrm{r,Al-NbSe_2}$) decreases with an increasing bias current (up to $I_\mathrm{DC} = \SI{30}{\micro\ampere}$), indicating a corresponding increase in kinetic inductance (Fig.~\ref{fig:measurement}e). This behavior is similar to the microwave response due to the Kerr-type nonlinearity. In the Ginzburg-Landau framework, the dependence of kinetic inductance of the superconducting film with bias current is well-described by a quadratic function (ignoring the terms beyond 2\textsuperscript{nd} order)~\cite{tinkham2004introduction}:
\begin{align}
    L_k (I_\mathrm{DC}) = L_k(0) \left[1+ \left(\frac{I_\mathrm{DC}}{I^\ast}\right)^2+...\right],
    \label{eq:Lk_current}
\end{align}
\noindent where $L_\mathrm{k}(0)$ is the kinetic inductance of the resonator at zero current bias, and $I^{\ast}$ is a characteristic current, also known as the depairing current. $I^{\ast}$ is typically comparable to the superconducting critical current $I_\mathrm{c}$, and sets the scale of non-linearity. By fitting to this relationship, $I^{\ast}$ is determined to be \SI{0.4}{\mA} for the \SI{9}{\nm} thick, $\SI{3.9}{\mu m}$ wide \ch{NbSe2} sample. We observe that both the resonant frequency and the kinetic inductance per square exhibit a quadratic dependence on the DC bias current (Fig.~\ref{fig:measurement}f), consistent with the anisotropic nodeless pairing of \ch{NbSe2} reported in previous studies~\cite{dvir2018spectroscopy,hamill2021two}. 

\subsection*{Layer Dependence of $L_\mathrm{k}$ in \ch{NbSe2}}
VdW superconductors remain superconducting in the monolayer limit \cite{xi2016ising,ugeda2016characterization}, which enables one to increase kinetic inductance by reducing the thickness. Figure~\ref{fig:Lk_layer}a shows the sheet kinetic inductance ($L_\mathrm{k,sq}$) of \ch{NbSe2} films extracted from microwave measurements as a function of thickness ($d$) and the corresponding layer numbers. The measured $L_\mathrm{k,sq}$ (red circles) exhibits an approximate $1/d$ dependence, as expected for thin superconducting films, and reaches \SI{1.2}{\nano\henry} for the monolayer sample (see Supplementary Information for the determination of layer numbers). The values are compared to those of other superinductors based on amorphous materials, including \ch{TiN}\cite{shearrow2018atomic, coumou2012microwave}, \ch{NbTiN}\cite{ samkharadze2016high, bretz2022high}, granular aluminum (gr-Al)\cite{winkel2020implementation}, and \ch{NbN}\cite{frasca2023nbn} across a range of film thicknesses up to \SI{13}{nm}, as shown in Fig.~\ref{fig:Lk_layer}b. Notably, \ch{NbSe2}, particularly in the few-monolayer regime, exhibits one of the highest sheet kinetic inductances reported at comparable thicknesses, surpassing those of many conventional high-impedance superconductors.

The observed layer dependence may provide insight into the underlying mechanisms responsible for the high $L_\mathrm{k}$, including the role of scattering processes, and whether our \ch{NbSe2} devices operate in the dirty limit, the clean limit, or an intermediate regime. We consider a general expression for the sheet kinetic inductance $L_\mathrm{k,sq}$ that spans from the clean to the dirty limit (see Supplementary Information for the derivation):
\begin{equation}
    L_\mathrm{k,sq} = \frac{1}{d}\left(\frac{m_\mathrm{eff}}{n_0 e^2} + \frac{\hbar}{1.764 k_\mathrm{B}} \frac{\rho_\mathrm{s}}{T_\mathrm{c}}\right) = 
    \frac{m_\mathrm{eff}}{n_0 e^2 d} + \frac{\hbar}{1.764 k_\mathrm{B}} \frac{R_\mathrm{s}}{T_\mathrm{c}}
    \label{eq:LK_clean_dirty_maintext}
\end{equation}
% \begin{equation}
%         L_\mathrm{k,dirty} = \frac{R_\mathrm{s}h}{2\pi^2 \Delta} \frac{1}{\tanh\left(\frac{\Delta}{2k_\mathrm{B}T}\right)} \left(\frac{l}{w}\right), 
%        \label{eq:LK_estimation_main}
% \end{equation}
Here, $\rho_\mathrm{s}$ denotes the normal-state sheet resistivity, $\hbar$ is the reduced Planck’s constant, $k_\mathrm{B}$ is the Boltzmann constant, and $R_\mathrm{s}$ is the normal-state sheet resistance. If we assume that the term in parentheses varies slowly with film thickness, then we expect the approximate $1/d$ dependence that we see in Fig.~\ref{fig:Lk_layer}a. This expression suggests that the total kinetic inductance arises from a combination of clean-limit (first term) and dirty-limit (second term) contributions. 
%Consequently, the clean-limit contribution to the kinetic inductance, given by $\tfrac{m_\mathrm{eff}}{n_0 d e^2}$, is thickness-independent. 
Figure~\ref{fig:Lk_layer}c plots the measured sheet kinetic inductance $L_\mathrm{k,sq}$ as a function of $R_\mathrm{s}/T_c$. The $R_\mathrm{s}$ and $T_\mathrm{c}$ values are obtained from both our transport measurements and previous studies (shown in green and black circles in Fig.~\ref{fig:Lk_layer}c, summarized in Table~\ref{table:SI3}). We observe a linear dependence for the larger values of $L_\mathrm{k,sq}$ in Fig.~\ref{fig:Lk_layer}c. A linear fit yields a slope of \SI{16.3}{} which characterizes the dirty-limit dependence, and the y-intercept of \SI{80.0}{pH} shows that there is a significant contribution from the clean-limit for thicker samples (note that the $1/d$ dependence for the clean-limit term can be considered, but qualitatively, the contribution from the clean-limit term remains significant). As the sample thickness approaches the monolayer limit, the ratio of the measured kinetic inductance to the clean-limit kinetic inductance (${L_{k,sq}}/{L_{k,sq,clean}}$) increases from 0.7 to 15 monotonically, indicating that the dirty-limit contribution becomes increasingly significant in thinner \ch{NbSe2} samples. This model captures the crossover from clean to the dirty-limit behavior as the sample thickness decreases.  

% A linear fit yields a slope of \SI{16.3}{} and a y-intercept of \SI{80.0}{pH}, representing the clean-limit contribution to $L_\mathrm{k,sq}$. 
% %Reduced dimensionality may introduce extra contributions that increase the kinetic inductance.

We now discuss possible origins of the observed crossover behavior. In thicker \ch{NbSe2} films, the majority of charge carriers reside in the interior layers of the 2D crystal. The clean-limit-like behavior observed in this regime—corresponding to the lower-left corner of Fig.~\ref{fig:Lk_layer}c—may be attributed to the high crystallinity of the bulk material and effective screening from carriers in the top and bottom \ch{NbSe2} layers. In contrast, thinner devices exhibit a greater influence from their surrounding environment. In the monolayer limit, all charge carriers are in close proximity to the encapsulating hBN on both sides, leading to enhanced surface scattering from impurities or fabrication residues. This increased scattering may result in more pronounced dirty-limit contributions to the kinetic inductance. Moreover, \ch{NbSe_2} has been shown to host two superconducting bands with distinct gap sizes, with the larger gap decreasing monotonically as thickness is reduced~\cite{dvir2018spectroscopy}. In other multi-band superconductors~\cite{homes2015fete,dai2016coexistence}, such disparity across the Fermi surface has been linked to the coexistence of clean- and dirty-limit behaviors, arising from band-dependent coherence lengths and scattering rates. Our experiment does not explicitly resolve contributions from individual bands, but similar multi-band effects may contribute to the observed layer dependence of kinetic inductance in \ch{NbSe2} (Fig.~\ref{fig:Lk_layer}c). Additionally, as the superconducting gap (or $T_\mathrm{c}$) decreases with reduced thickness, the superconducting coherence length becomes longer. For a fixed mean free path, this leads to a larger coherence length to mean free path ratio, indicating a shift toward dirty-limit behavior in the thinner samples. We note that the combined contributions from both clean- and dirty-limit superconductivity observed in the microwave measurements here contrast with previous DC transport studies, which reported only clean-limit behavior. This discrepancy suggests that the microwave response may be sensitive to the multi-band nature of \ch{NbSe2}~\cite{prober1980upper, khestanova2018unusual, zhu2022transport, benyamini2019blockade}. 

%Moreover, it has been shown that \ch{NbSe_2} exhibits two superconducting bands with distinct gap sizes, and the larger gap reduces monotonically as the thickness decreases \cite{dvir2018spectroscopy}. As observed in other multi-band superconductors~\cite{homes2015fete,dai2016coexistence}, this characteristic could result in simultaneous clean- and dirty-limit behavior, as differences in respective Fermi surface leads to varying coherence lengths and scattering rates. Additionally, as the thickness decreases, the coherence length $\xi_0$ increases for a given mean free path ($l_\mathrm{mfp}$), leading to a higher $\xi_0 / l_\mathrm{mfp}$ ratio, thereby indicating stronger contributions from dirty-limit behavior.

% Joel's edit:

Finally, in thin superconducting films where the thickness $d$ is much smaller than the penetration depth $\lambda$, the relevant length scale becomes the Pearl length $\Lambda = \frac{2\lambda^2}{d}$~\cite{pearl1964current}.  All of our \ch{NbSe2} samples operate in this Pearl regime ($d \ll \lambda \sim \SI{230}{nm}$~\cite{fridman2025anomalous}), with both thickness and width much smaller than the estimated Pearl length (10-\SI{70}{\um}; see Supplementary Information). Recent studies on high-$T_{\mathrm{c}}$ YBCO superconductors have shown that geometric confinement in this regime can strongly enhance the kinetic inductance as the device width approaches or falls below $\Lambda$ \cite{srivastava2025}. While a detailed investigation of this effect in \ch{NbSe2} lies beyond the scope of this work, our observation of large $L_\mathrm{k}$ values ---  and a steeper-than-expected inverse thickness scaling, relative to the prediction of Eq.~\ref{eq:LK_clean_dirty_maintext} --- may be qualitatively consistent with such confinement-enhanced kinetic inductance in the Pearl regime.

\section*{Conclusions}
We characterize the kinetic inductance and nonlinear properties of \ch{NbSe2} using a coplanar waveguide resonator terminated by hBN-encapsulated \ch{NbSe2} thin films. %The \ch{NbSe2} device dimensions lie within the Pearl regime ($d < \lambda$), with both thickness and width well below the Pearl length $P_{\mathrm{L}}$. 
The kinetic inductance shows an approximate $1/d$-type scaling for the thinner samples and reaches \SI{1.2}{\nano \henry}/$\Box$ at monolayer thickness --- among the highest reported for superconductors within this thickness regime.
%We also show that the kinetic inductance exceeds values estimated from the normal-state resistance by at least one order of magnitude, deviating from the behavior expected in dirty superconductors. These results suggest that the observed kinetic inductance is enhanced by reduced dimensions, rather than by disorder-induced scattering as in amorphous, dirty superconductors. 
Under the microwave drive, the device exhibits a nonlinear frequency shift that is consistent with the nonlinear Kerr model. The extracted Kerr coefficients ($K/2\pi$ = –0.008 to 14.7 Hz/photon) are relatively small, making \ch{NbSe2} well-suited for applications that require large linear inductance with minimal nonlinearity, such as the linear shunt-inductor used in fluxonium qubits and in photon detectors operating in the microwave regime~\cite{shein2024fundamental}. The $L_\mathrm{k}$ exhibits a quadratic dependence on DC bias current, consistent with the Ginzburg-Landau framework.

The measured kinetic inductance in \ch{NbSe2} exhibits a crossover from clean- to dirty-limit behavior as the thickness decreases. This transition can be qualitatively described using a generalized expression for $L_\mathrm{k}$ derived from BCS theory. In the ultra-thin regime, enhanced contributions from the dirty limit may arise due to increased surface scattering, multi-band superconductivity, and geometric confinement.

We demonstrate that \ch{NbSe2} is a crystalline two-dimensional superconductor that exhibits large, linear inductance. The hBN-\ch{NbSe2}-hBN heterostructure is well suited for compact lump-element devices, such as shunt inductors for fluxonium qubits or readout resonators, while minimizing parasitic coupling. These results also highlight the ability of microwave measurements to probe superconducting properties in low-dimensional materials, offering a valuable complement to DC transport and other techniques. 

%The fabrication and characterization techniques demonstrated here can be extended to a broad range of two-dimensional superconductors to explore novel functionalities for applications and to investigate their fundamental properties. 

%These results highlight the potential of two-dimensional crystalline superconductors to achieve large kinetic inductance with minimal dissipation. %Further investigation of the intrinsic quality factor $Q_\mathrm{i}$ will inform the development of low-loss superconducting quantum circuit elements and sensors for next-generation quantum technologies using van der Waals superconductors and heterostructures.

\section*{Acknowledgements}
The authors thank Shoumik D. Chowdhury, Aziza Almanakly, Lamia Ateshian, David Rower, Xirui Wang, Xueqiao Wang, and the device packaging team at MIT Lincoln Laboratory for their assistance in measurement, fabrication, and packaging. This work was carried out in part using the MIT.nano's facilities. This research was funded in part by the US Army Research Office grant no.~W911NF-22-1-0023, by the  National Science Foundation grant no. 2412810 and no.~OMA-1936263, by the Air Force Office of Scientific Research grant no.~FA2386-21-1-4058, and under Air Force Contract No.~FA8702-15-D-0001. S.Z. acknowledges support from the Lisa Su (1990) Fellowship awarded by the EECS Department at MIT  as well as the Faculty for the Future Fellowship from the Schlumberger Foundation. T. W. and P. K. acknowledge support from DOE contract DE-SC0012260. M.H. is supported by an appointment to the Intelligence Community Postdoctoral Research Fellowship Program at the Massachusetts Institute of Technology, administered by Oak Ridge Institute for Science and Education (ORISE) through an interagency agreement between the U.S. Department of Energy and the Office of the Director of National Intelligence (ODNI). D. R-L. acknowledges support from the Rafael del Pino Foundation. R.C. and A.K.D. acknowledge support from the U.S. Department of Energy, Office of Science National Quantum Information Science Research Center's Co-design Center for Quantum Advantage (C2QA) under contract number DE-SC0012704. R.C. and A.K.D. contributed to this work by performing low-frequency Raman spectroscopy measurements to determine the layer number/thickness of thin NbSe$_2$ flakes from shear-mode frequency analysis, and this contribution was supported by C2QA. K.W. and T.T. acknowledge support from the JSPS KAKENHI (Grant Numbers 21H05233 and 23H02052), the CREST (JPMJCR24A5), JST, and the World Premier International Research Center Initiative (WPI), MEXT, Japan. P.J-H. acknowledges support by the Air Force Office of Scientific Research (AFOSR) grant FA9550-21-1-0319, the Office of Naval Research (ONR) grant N000142412440, the MIT/Microsystem Technology
Laboratories, Samsung Semiconductor Research Fund, the Gordon and Betty Moore Foundation’s EPiQS Initiative through Grant No. GBMF9463, the Fundacion Ramon Areces, and the CIFAR Quantum Materials program. Any opinions, findings, conclusions, or recommendations expressed in this material are those of the author(s) and do not necessarily reflect the views of the U.S. Air Force or the U.S. Government.

\section*{Author Contributions}
J.\^I-j.W. conceived and designed the experiment. S.Z., J.\^I-j.W, and T.H.D. performed the microwave simulations. S.Z., T. W., M.T., J.\^I-j.W., D. R-L., A.G., A.K.D., D.K.K., and B.M.N. contributed to the device fabrication and characterization. S.Z., J.\^I-j.W., T. W., R. A., M. H., and D. R-L participated in the measurements. S.Z., J.\^I-j.W., R. A., and M.H. analyzed the data. K.W. and T.T. grew the hBN crystal. S.Z., J.\^I-j.W., and W.D.O. led the paper writing, and all other authors contributed to the text. J.\^I-j.W., J.A.G., K.S. M.E.S., R.C., P.K., P.J-H., T.P.O, and W.D.O supervised the project.    

\section*{Competing Interests Statement}
The authors declare no competing interests.
\pagebreak
\pagebreak

\begin{table}[H]
\centering

\renewcommand{\arraystretch}{2}
\begin{tabular}{ | >{\centering}p{3em} | >{\centering}p{2.5em} | >{\centering}p{2.5em} | >{\centering}p{2.5em} | >{\centering}p{5em} | >{\centering}p{3.8em} | >{\centering}p{3em} | >{\centering}p{3.4em} | >{\centering}p{5.2em} |  >{\centering\arraybackslash}p{3.5em} |}
\hline
\textbf{Device ID} & \textbf{d \\(\SI{}{\nano\meter})} & \textbf{W (\SI{}{\micro\meter}) } & \textbf{L} (\SI{}{\micro\meter}) & \textbf{$\bm{\textit{f}_\mathrm{r,Al-NbSe_2}}$ (\SI{}{\giga\hertz})} & \textbf{$\bm{\textit{f}_\mathrm{r,Al}}$ (\SI{}{\giga\hertz})} &\textbf{$\bm{\textit{L}_\mathrm{k}}$ (\SI{}{\nano\henry})} & \textbf{$\bm{\textit{L}_\mathrm{k,sq}}$ (\SI{}{\nano\henry/sq})} & \textbf{$\bm{\textit{Q}_\mathrm{i,Al-NbSe_2}}$} &\textbf{$\bm{\textit{K}/2\pi}$ (\SI{}{\hertz})} \\ \hline
D1 & 1.99 & 3.1 & 3.1 & 3.62 & 5.30 & 1.195 & 1.195 & 0.8$\times10^4$ & -14.7\\  \hline
D2 & 4.08 & 2.5 & 5.7 & 4.14 & 5.21 & 0.643 & 0.282 & 6.0$\times10^4$  & -3.8\\  \hline
D3 & 5.86 & 2.6 & 3.0 & 4.88 & 5.33 & 0.220 & 0.191 & 0.7$\times10^4$ & -2.9 \\  \hline 
D4 & 7.13 & 4.8 & 6.7 & 4.93 & 5.35 & 0.200 & 0.143 & 2.4$\times10^4$ & -2.6 \\  \hline 
D5 & 7.98 & 4.5 & 12.0 & 4.63 & 5.33 & 0.360 & 0.135 & 1.2$\times10^4$ & -1.9\\  \hline 
D6 & 8.81 & 3.9 & 8.6 & 4.79 & 5.34 & 0.274 & 0.124 & 2.8$\times10^4$  & -1.7\\  \hline
D7 & 8.90 & 4.5 & 4.2 & 5.26 & 5.49 & 0.100 & 0.107 & 2.1$\times10^4$ & -1.5\\  \hline
D8 & 10.07 & 2.5 & 6.6 & 5.16 & 5.49 & 0.149 & 0.056 & 0.7$\times10^4$ & -0.4\\  \hline
D9 & 11.89 & 6.0 & 6.9 & 5.46 & 5.60 & 0.057 & 0.049 & 0.6$\times10^4$ & -0.03\\  \hline
D10 & 24.73 & 4.1 & 7.5 & 5.34 & 5.37 & 0.014 & 0.008 & 18$\times10^4$ & -0.006\\  \hline
D11 & 30.20 & 3.4 & 6.1 & 5.42 & 5.44 & 0.009 & 0.005 & 2.5$\times10^4$ & -0.006\\  \hline
D12 & 52.12 & 2.3 & 4.6 & 5.6 & 5.62 & 0.008 & 0.004 & 5.8$\times10^4$ & -0.008\\  \hline
\end{tabular}
\caption{\textbf{Summary of kinetic microwave characterization of \ch{NbSe2} samples across different thicknesses.} W and L represent the nominal width and nominal length of the \ch{NbSe2} samples with a thickness of d. $\textit{Q}_\mathrm{i,Al-NbSe_2}$ and Kerr coefficients ($K/2\pi$) are extracted from \ch{NbSe2}-terminated Al resonators. The film thickness ($d$) is measured with atomic force microscopy (AFM).}

\end{table}

\pagebreak

\pagebreak

\pagebreak

%\begin{figure}[hbtp]
\begin{figure}[H]
    \centering
    \includegraphics[width= \textwidth]{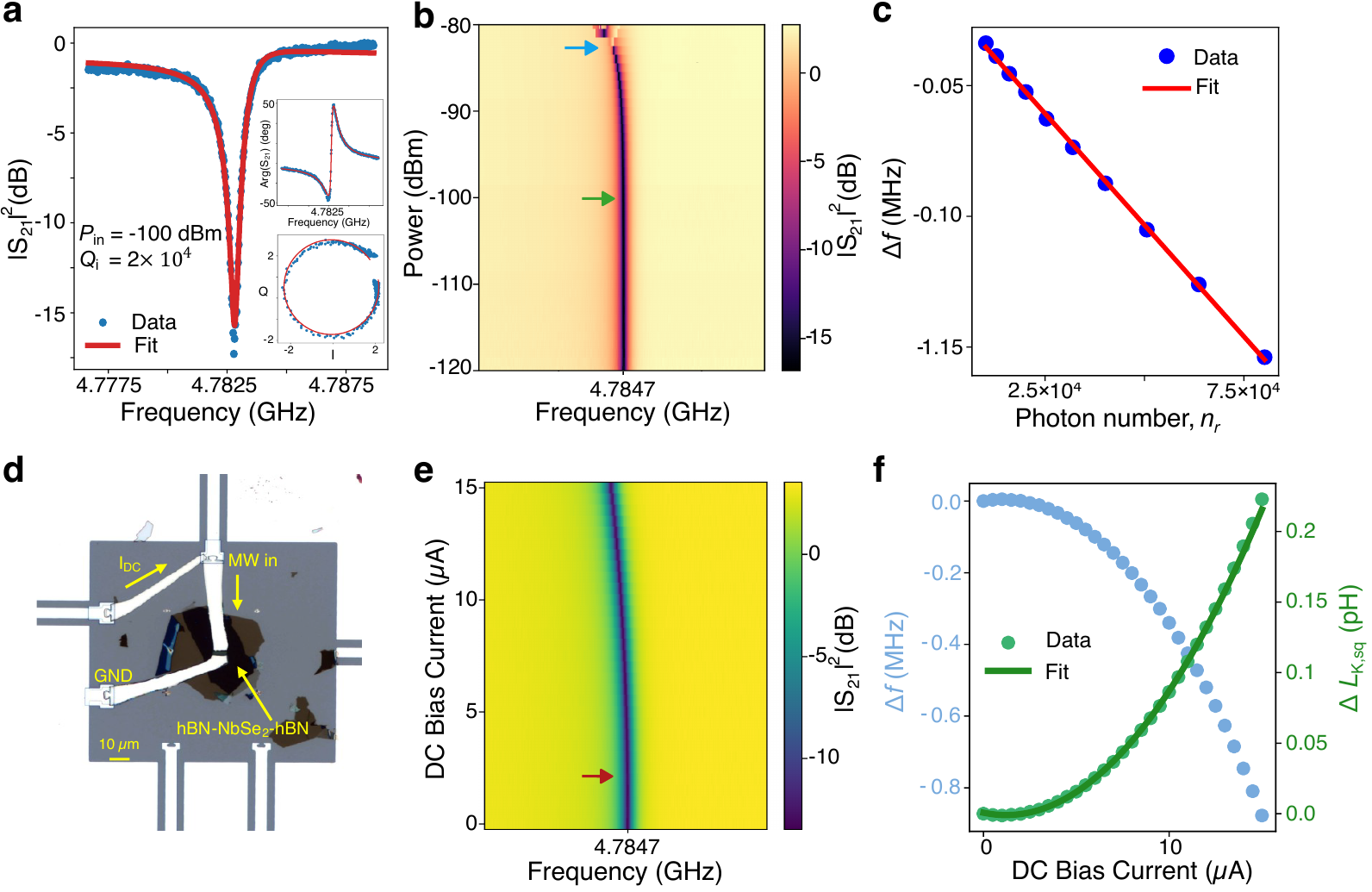}
    \caption{\textbf{Microwave measurements of \ch{NbSe2}-terminated $\lambda$/4-resonator.} \textbf{a,} Transmission coefficient ($|S_{21}|$) of a \ch{NbSe2}-terminated $\lambda$/4-resonator measured at \SI{10}{mK} (Device ID D6). Inset: Complex I–Q plane showing the measured response (blue) and Lorentzian fit (red). \textbf{b,} Transmission coefficient ($|S_{21}|$) of \ch{NbSe2}-terminated $\lambda$/4-resonator resonator as a function of input microwave power ($P_{rf}$). The resonant frequency ($f_\mathrm{r,Al-NbSe_2}$) shifts to the lower frequency with increasing microwave power. The onset of the frequency shift---defined as exceeding one standard deviation from the low-power baseline ($f_\mathrm{r,Al-NbSe_2} =$ \SI{4.785}{GHz})---is observed at an input power of \SI{-100}{dBm} (indicated by the green arrow). At \SI{-82}{dBm}, the resonant peak bifurcates (blue arrow), indicating the onset of a bistable regime. \textbf{c,} Linear fit of the resonant frequency shift ($\Delta f = f_\mathrm{r,Al-NbSe_2}(n_\mathrm{r}= 1) - f_\mathrm{r,Al-NbSe_2}(n_\mathrm{r}$)) as a function of resonator photon number ($n_\mathrm{r}$) in the \ch{NbSe2}-terminated $\lambda$/4-resonator. \textbf{d,} Optical image of the \ch{NbSe2}-terminated $\lambda$/4-resonator with a DC bias line connected at the microwave input port. \textbf{e,} DC bias dependence of the \ch{NbSe2}-terminated $\lambda$/4-resonator measured at a fixed microwave power of $P_{rf} =$ \SI{-120}{dBm}. The resonant frequency shifts to the lower frequency with increasing bias current. The onset of the frequency shift—defined as exceeding one standard deviation from the baseline ($f_\mathrm{r,Al-NbSe_2} =$ \SI{4.785}{GHz})—is observed at a bias current of $\SI{1}{\mu A}$ (maroon arrow). \textbf{f,} Extracted resonant frequency shift ($\Delta f = f_\mathrm{r,Al-NbSe_2}(I_\mathrm{DC}= 0) - f_\mathrm{r,Al-NbSe_2}(I_\mathrm{DC})$) and kinetic inductance shift ($\Delta L_\mathrm{k,sq} = L_\mathrm{k,sq}(I_\mathrm{DC}= 0) - L_\mathrm{k,sq}(I_\mathrm{DC})$) of the \ch{NbSe2}-terminated resonator as a function of DC bias current. Measured data (circle-green and sky-blue) and theoretical fit (line-green) are shown.}
    \label{fig:measurement}
\end{figure}
\pagebreak

\pagebreak
\begin{figure}[H]
    \centering
    \includegraphics[width= \textwidth]{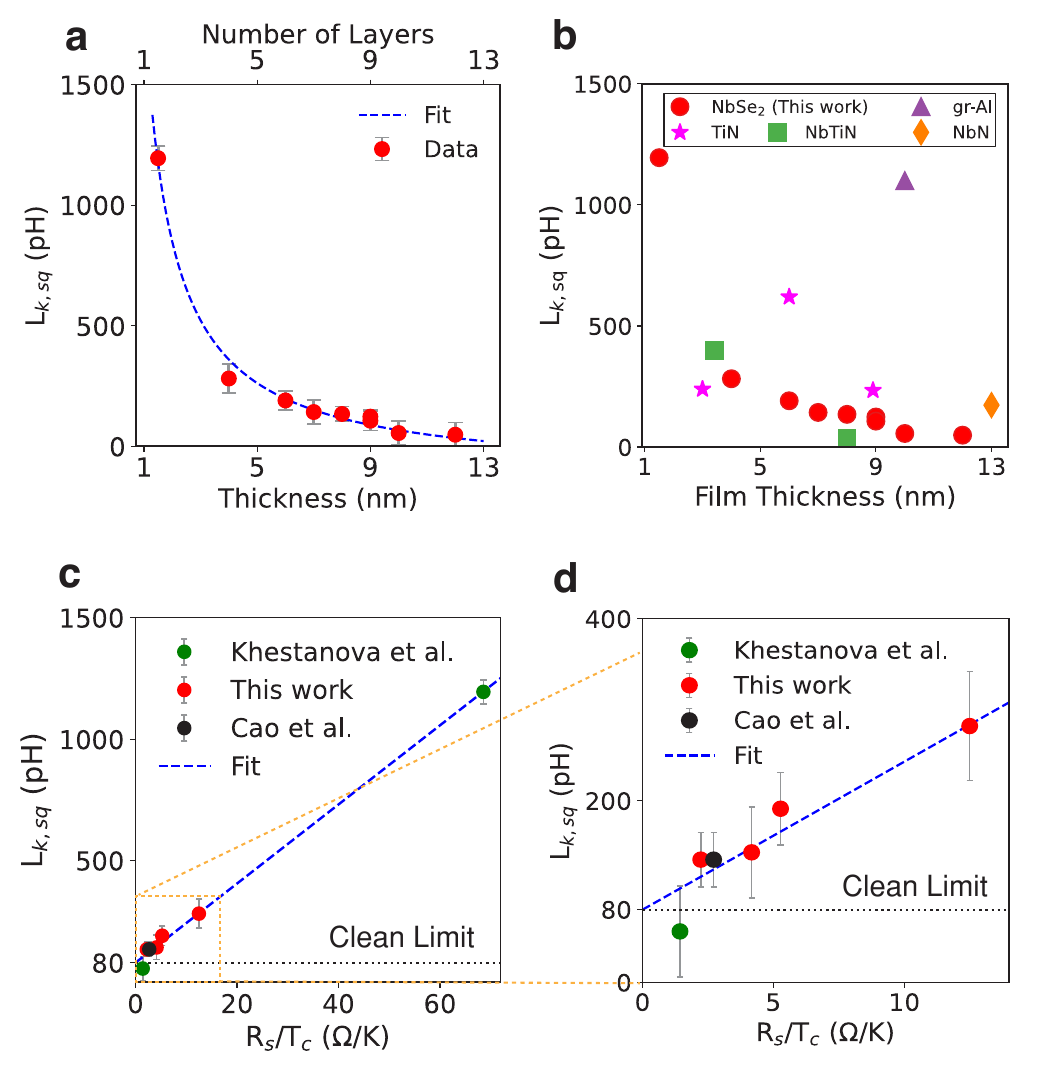}
    \caption{\textbf{Thickness dependence of kinetic inductance in \ch{NbSe2}.} \textbf{a,} Sheet kinetic inductance ($L_\mathrm{k,sq}$) from microwave measurements (circle-red) of nine \ch{NbSe2} devices with varying film thicknesses. The blue dashed line shows a $1/d$ scaling fit. \textbf{b,} Comparison of measured $L_\mathrm{k,sq}$ in \ch{NbSe2} with values reported for other high-impedance superconducting thin films \cite{bretz2022high, coumou2012microwave, samkharadze2016high, shearrow2018atomic, frasca2023nbn, winkel2020implementation}. \textbf{c} Sheet kinetic inductance ($L_\mathrm{k,sq}$) from microwave measurements is plotted as a function of $R_\mathrm{s}/T_c$, where $R_\mathrm{s}$ and $T_\mathrm{c}$ values are obtained from both our transport measurements (red circles) and previous studies (green and black circles)\cite{khestanova2018unusual,cao2015quality}. The blue dashed line is a linear fit with a slope of \SI{16.3}{} and a y-intercept of \SI{80.0}{pH}, marked as the clean-limit contribution to $L_\mathrm{k,sq}$. \textbf{d,} The magnified view of the lower region for clarity.}
    
    % Sheet kinetic inductance of \ch{NbSe2} estimated from sheet resistance and critical temperature. Estimates are based on our own DC transport measurements (hexagon-red) and on $R_\mathrm{s}$ and $T_\mathrm{c}$ values reported in the literature (star-green \cite{khestanova2018unusual} and pink \cite{cao2015quality}). Inset: Difference between $L_\mathrm{k,sq}$ values obtained from microwave measurements and DC transport estimates.}
    \label{fig:Lk_layer}
\end{figure}

\pagebreak
\pagebreak

\section*{References}
\printbibliography[heading=none]

\section*{Data availability}
The data that support the findings of this study are available from the corresponding author upon reasonable request and with the cognizance of our US Government sponsors, who funded the work.

\newpage
\part*{Supplementary Information}
\renewcommand{\thefigure}{S\arabic{figure}}
\setcounter{figure}{0}
\renewcommand{\thetable}{S\arabic{table}}
\setcounter{table}{0}

\tableofcontents
\section{Device Assembly and Fabrication}
Aluminum CPW resonators are fabricated from a highly resistive 2-inch Si wafer coated with \SI{250}{nm} MBE-grown aluminum film. The resonator circuit is patterned via photolithography and wet etched using Transene Aluminum etchant. Afterwards, the 2-inch wafer is diced into \SI{5}{mm} by \SI{5}{mm} chips. The bulk crystal of \ch{NbSe2} is grown by HQ Graphene. The mechanical exfoliation and encapsulation of thin \ch{NbSe2} layers with hBN are carried out inside an Argon-filled glove box, maintaining oxygen and water levels below \SI{0.1}{ppm}. This prevents oxidation of \ch{NbSe2}, keeping the devices within the low-disorder regime. The hBN-NbSe$_2$-hBN heterostructures are transferred on the chip using dry polymer-based techniques \cite{wang2019coherent, Wang2013, pizzocchero2016hot}. After transferring the hBN-\ch{NbSe2}-hBN stacks onto the pre-patterned chips, the van der Waals heterostructures are taken out of the glove box. To process edge-contact, first the stack goes through reactive-ion etching with \ch{O2}/Ar/\ch{CHF3} to open edges of \ch{NbSe2}. The samples are immediately loaded into an e-beam evaporator to do in-situ ion-milling to clean the exposed edges. Aluminum is then deposited onto the cleaned \ch{NbSe2} edge at a tilt angle of 30\textdegree  with the substrate rotating at 60 rpm, ensuring uniform coverage \cite{sinko2021superconducting}. The bridging between the contact aluminum and the pre-fabricated aluminum resonator and ground plane is done by another in-situ ion mill and a normal aluminum deposition~\cite{grunhaupt2017argon}.

\section{Optimization of the Contact Resistance between hBN encapsulated \ch{NbSe2} and Al}\label{section:optimization}
It is critical to establish a low-resistance, transparent contact (with near-zero contact resistance) between hBN encapsulated \ch{NbSe2} and the aluminum electrode for achieving impedance matching and accurate resonant frequency measurements. A low contact resistance ($R_c$) is required because if it exceeds the characteristic impedance ($\SI{50}{\Omega}$ for our case), the resonator behaves like a $\lambda$/2-resonator instead of the desired $\lambda$/4 mode terminated with \ch{NbSe2} sample. This happens because the higher resistance causes the end of the resonator to act like an open circuit, disrupting the intended resonant behavior. Additionally, minimizing $R_c$ reduces dissipation and enables high-Q, high-accuracy measurements for characterizing the kinetic inductance and other superconducting properties of \ch{NbSe2}. 

\begin{figure}[htbp]
    \centering
    \includegraphics[width=0.8\textwidth]{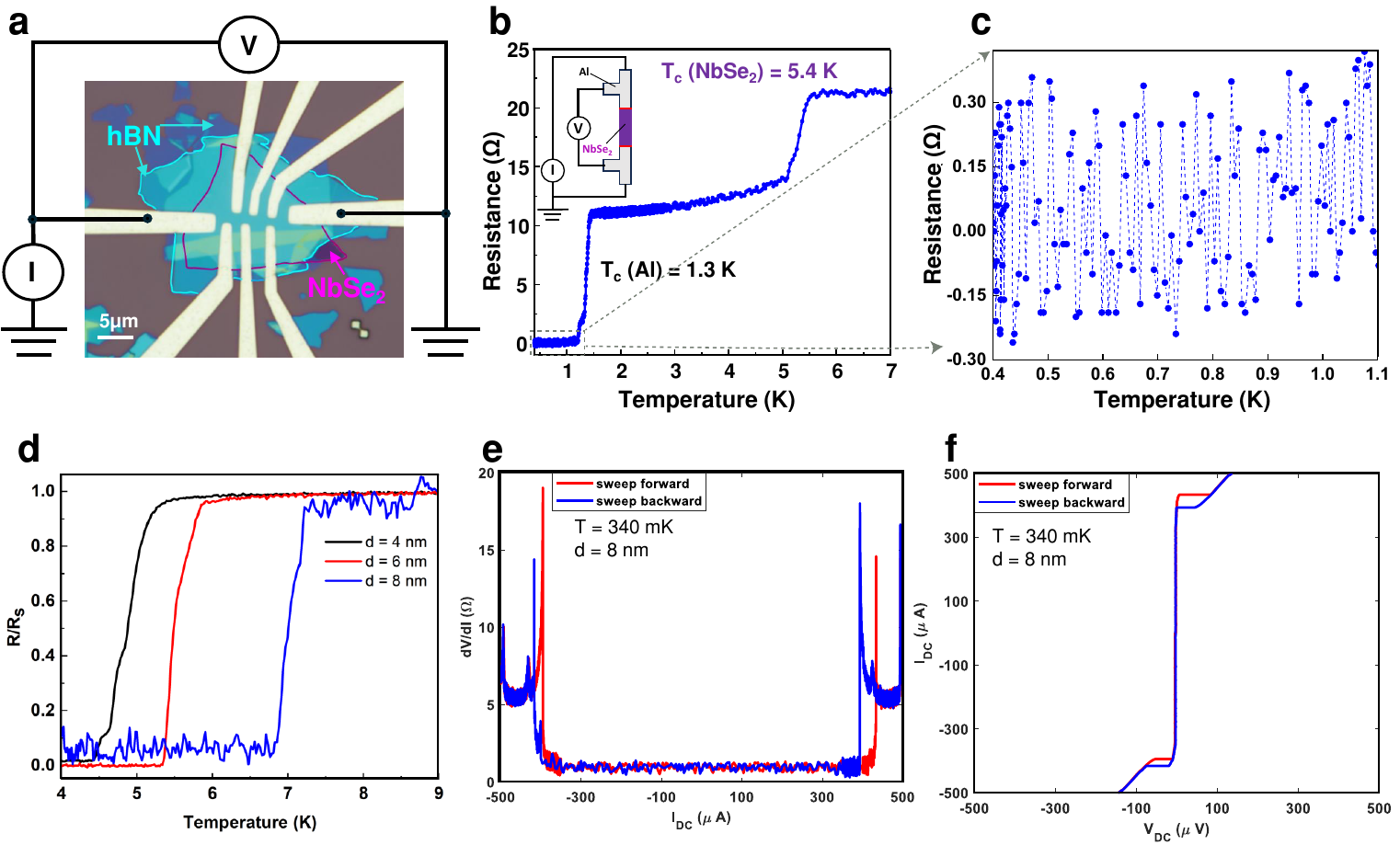}
    \caption{\textbf{Al and \ch{NbSe2} interface characterization and DC characteristics of \ch{NbSe2}.} \textbf{a,} Optical micrograph of a pseudo-4-probe DC device where hBN flakes fully encapsulate \ch{NbSe2} flake, the current is applied between two terminals, and voltage is measured across those two terminals without taking into account the line resistance of the measurement setup. \textbf{b,} Resistance(T) for Al-\ch{NbSe2}-Al device showing the superconducting transition of \ch{NbSe2} (d = 6 nm) at 5.4 K and the superconducting transition of Al at 1.3 K. Inset represents the 4-pt measurement setup for measuring the contact resistances of the Al/\ch{NbSe2} interfaces. \textbf{c,} Total Resistance of the device below Al transition, which has dropped to zero within the noise floor of the measurement. \textbf{d,} Temperature dependence of the resistance for devices with \ch{NbSe2} thicknesses of 4 nm, 6 nm, and 8 nm. To extract the $T_\mathrm{c}$, we use the mean-field definition of $T_\mathrm{c}$, which corresponds to half of the normal state resistance. \textbf{e,} Differential resistance (dV/dI) measurement for the d = 8 nm thick \ch{NbSe2} of 4-point DC measurement. The plot shows the critical current $\SI{400}{\mu A}$ for that device. \textbf{f,} Current-voltage (I-V) relation on the same device.}
    \label{fig:dc}
\end{figure}

We fabricate a few devices for DC measurement using the above fabrication procedures to confirm a reliable superconducting contact between thin \ch{NbSe2} and Al. We vary the ion-milling duration to determine the optimal time needed to remove oxides from the edges of \ch{NbSe2}. To eliminate the line resistances, we measure the devices in a pseudo-4-probe configuration, Fig.~\ref{fig:dc}a. Below the transitions of both \ch{NbSe2} and Al, the total resistance of the device, including the Al-\ch{NbSe2} junctions, drops to zero within the noise floor of our measurement ($R_\mathrm{c} = \SI{0.01}{\Omega}$) Fig.~\ref{fig:dc}b-c. We present the results of the measurement of DC contact devices to optimize the contact resistance as a function of the ion milling time. The results are summarized in the Table~\ref{table: SI1}. We chose the optimum ion milling time of 105 seconds for our microwave devices presented in the main text.
\begin{table}[h]
\centering
\begin{tabular}{ |c|c|c|c|c|c| }
\hline
$\textbf{Ion-milling time (sec)}$ & $\textbf{$\textit{R}_\mathrm{c}$ ($\Omega$/\um)}$ & $\textbf{$\textit{R}_\mathrm{s}$ ($\Omega$/sq)}$ & $\textbf{T$_\mathrm{c}$ (K)}$ & $\textbf{d (nm)}$ \\ \hline
120 & 0.01 & N/A & N/A & 10 \\  \hline
105 & 0.01 & 31 & 5.5 & 6 \\  \hline 
90 & 0.01 & 19 & 7 & 8 \\  \hline 
90 & 0.02 & N/A & N/A & 4 \\  \hline 
80 & 1,80 & 60 & 4.8 & 4 \\  \hline 
75 & 0.02,2 & 29 & 5.5 & 6 \\  \hline 
60 & 0.02 & 45 & 4.9 & 5 \\  \hline  
\end{tabular}
\caption{\textbf{Summary of DC samples for contact resistance characterization.} The in-situ ion-milling time and the thickness of \ch{NbSe2} both have been varied to characterize the contact resistance between \ch{NbSe2} edge and aluminum interface.}
\label{table: SI1}
\end{table}
The bulk \ch{NbSe2} undergoes a transition into the charge-order phase at T $\sim$ \SI{33}{K}, and a transition into the superconducting phase at T $\sim$ \SI{7}{K} \cite{hamill2021two,sanna2022real, cao2015quality}. As the thickness of \ch{NbSe2} decreases, the critical temperature ($T_\mathrm{c}$) decreases, as shown in Fig.~\ref{fig:dc}d, our results are consistent with other studies in the literature \cite{khestanova2018unusual}. All our \ch{NbSe2} samples are high-quality crystals with relatively high residual resistance ratios ($R_\mathrm{300K}/R_\mathrm{N}$ = 5-10) \cite{hamill2021two}, the normal-state sheet resistance ($R_\mathrm{s}$) right above the superconducting transition increases as the thickness of \ch{NbSe2} decreases; the values are summarized in Table~\ref{table: SI1}. Figure~\ref{fig:dc}e-f shows four-terminal transport characterization of one of the devices (d = 8 nm) using current biasing. The differential resistance was measured by applying a small AC excitation current (\SI{1}{nA}) superimposed on a DC bias current and detecting the AC voltage (dV) between two electrodes. The critical current of the \ch{NbSe2} sample is of order \SI{0.4}{mA}, which is consistent with the previous study \cite{shein2024fundamental, yokoi2020negative}.

\section{Measurement Setup}
\begin{figure}[htbp]
    \centering
    \includegraphics[width=0.8\textwidth]{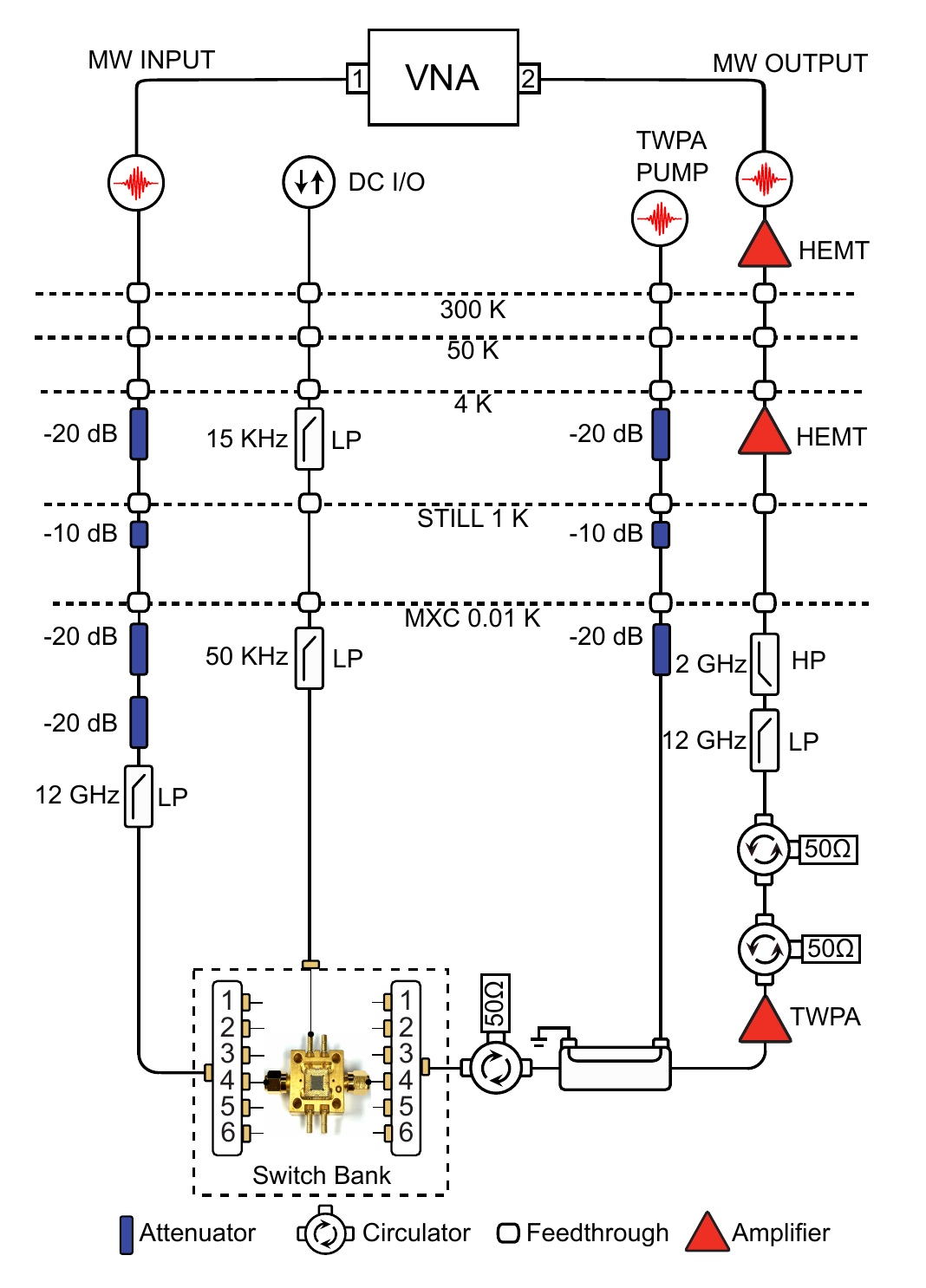}
    \caption{\textbf{Measurement setup.} Wiring diagram of the microwave and DC setup to the device.}
    \label{fig:meas_chain}
\end{figure}

Figure~\ref{fig:meas_chain} shows the wiring diagram of the measurement setup used for this work. To send the microwave signals down to the package, we use semi-rigid coaxial cables with low thermal conductivity and low-loss superconducting niobium-titanium (NbTi) cables for the readout. Within our readout input lines, we have integrated attenuation mechanisms: a \SI{20}{dB} attenuator at \SI{4}{K}, a \SI{20}{dB} attenuator at the still plate, and a \SI{30}{dB} attenuator at the mixing chamber stage. The output signals pass through a \SI{2}{GHz} high-pass filter and a \SI{12}{GHz} low-pass filter. We have a Josephson traveling-wave parametric amplifier (TWPA) in our read-out chain, and the TWPA pump tone is combined with the readout signal using a directional coupler, and the signal undergoes a final amplification stage employing a low-noise high-electron-mobility transistor (HEMT) amplifier positioned at the \SI{4}{K} stage and a room temperature amplifier. DC lines connected to the device are filtered at \SI{4}{K} with an RC $\pi$-filter and an RF+RC filter with a cut-off frequency of \SI{50}{kHz} at the mixing chamber stage. The control electronics used in this experiment are listed in the Table~\ref{table:SI2}.

\begin{table}[h]
\centering
\singlespacing
\begin{tabular}{c c c}
\hline\hline
Component & Manufacturer & Model \\ [0.5ex] 
\hline
Dilution Refrigerator & Bluefors & XLD1000 \\ 
VNA & Keysight & N5232A \\
DC Source & QDevil & QDAC \\
Lock-in Amplifier & Stanford Research Systems & SR830 \\
Multimeter & Keithley & 2010 \\ [1ex] 
\hline\hline
\end{tabular}
\caption{\textbf{Summary of control electronics.} Specifications of the control equipment used in the experiment, with their manufacturer and Model numbers.}
\label{table:SI2}
\end{table}

\section{Resonator Circular Fit}
\begin{figure}[htbp]
    \centering
    \includegraphics[width=0.8\textwidth]{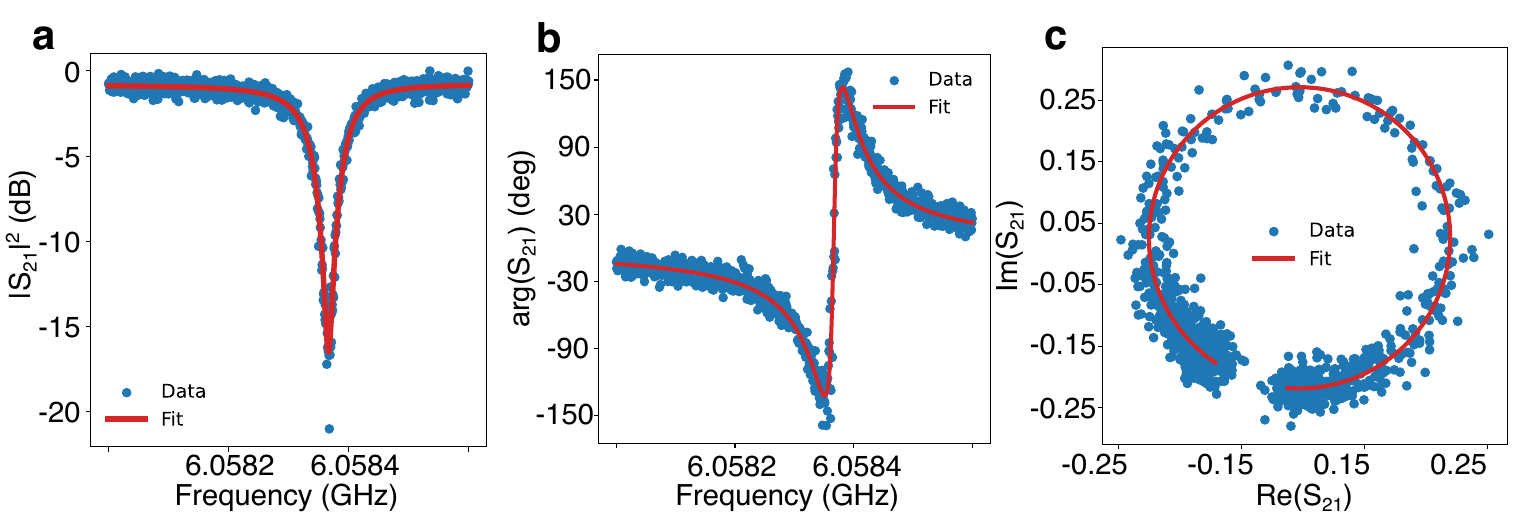}
    \caption{\textbf{Raw data (blue) and Lorentzian fit (red) of Al ``spectator” $\lambda$/4-resonator.}\textbf{a-b} Magnitude and phase response of $S_{21}$ signal as a function of frequency at the single photon limit. The equation \ref{eq:res_fit} function is used to fit the complex response. \textbf{c,} The imaginary part of $S_{21}$ is plotted versus its real part, where the circle diameter corresponds to $Q_\mathrm{L}/Q_\mathrm{c} = \frac{Q_\mathrm{i}/Q_\mathrm{c}}{(Q_\mathrm{i}/Q_\mathrm{c}) +1}$, which determines the contributions of coupling and internal losses of the resonator. The fitted resonance frequency is 6.0583 GHz, with an internal quality factor, $Q_\mathrm{i}$ is $5.985 \times 10^6$, and a coupling quality factor, $Q_\mathrm{c}$ is $10^5$. }
    \label{fig:meas_Al_shunted_control}
\end{figure}

For notch-type geometry, where the resonator is coupled to a transmission line, the frequency-dependent transmission $S_{21}$ is given by~\cite{Probst2015,baity2024circle}: 
\begin{equation}
    S_{21}^{notch} \left(f\right) = a e^{i\alpha} e^{-2\pi i f \tau} \left[1- \frac{\left(\frac{Q_\mathrm{L}}{|Q_\mathrm{c}|}\right)e^{i\phi}}{1+2iQ_\mathrm{L}\left(\frac{f}{f_\mathrm{r}}-1\right)}\right]
    \label{eq:res_fit}
\end{equation}
where $f$ is the signal frequency, $f_\mathrm{r}$ is the resonance frequency, $Q_\mathrm{L}$ is the loaded quality factor, $Q_\mathrm{c}$ is the coupling quality factor, and $\phi$ corresponds to the phase shift of resonance due to the impedance mismatch with the transmission line. The prefactor ($a e^{i\alpha}$) includes the overall gain and phase delay of the transmission line, including all the attenuation and amplification. Additionally, $\tau$ denotes the propagation time in the transmission line, which introduces a phase shift proportional to the frequency $f$. Figure~\ref{fig:meas_Al_shunted_control} shows the raw data (blue) and the Lorentzian fit using Eq.~\ref{eq:res_fit} (red) of the spectator resonator in the single-photon limit. We extract the values of the loaded quality factor ($Q_\mathrm{L}$) and the coupling quality factor ($Q_\mathrm{c}$) from the fit. The internal loss determines the intrinsic quality ($Q_\mathrm{i}$) of the resonator extracted from: $\frac{1}{Q_\mathrm{L}} = \frac{1}{Q_\mathrm{i}} + \frac{1}{Q_\mathrm{c}}$. The fitted internal quality factors ($Q_\mathrm{i}$) of six different $\lambda/4$-resonators are plotted in Fig.~\ref{fig:meas_Al_shunted}c. 

\section{Non-linearity of the Resonator from the Inductance}

\begin{figure}[htbp]
    \centering
    \includegraphics[width=0.8\textwidth]{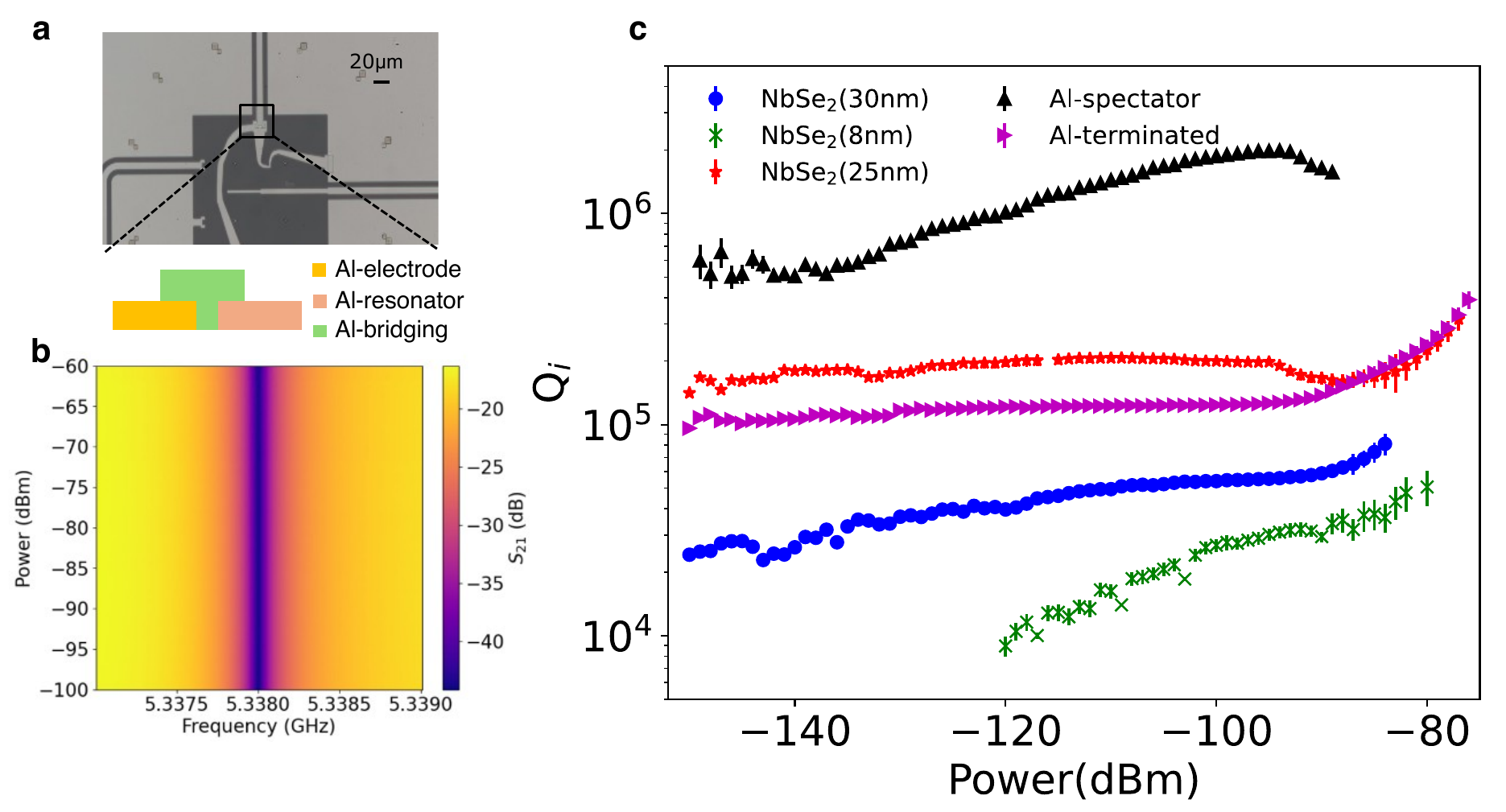}
    \caption{\textbf{Measurements of $\lambda$/4 resonator.} \textbf{a,} Optical image of Al-terminated resonator of similar geometry used in Fig 3 (main text). Here, two different Al layers (one MBE grown (orange) and another e-beam evaporated (yellow)) are made contact through an aluminum bridging step (green). The bridging step is performed by doing in-situ ion-milling to remove the native oxide layer and evaporate another layer of aluminum. \textbf{b,} Magnitude response of Al-shunted resonator with varying the input microwave power from \SI{-100}{dBm} to \SI{-60}{dBm}. The resonant frequency has a very negligible change, showing a small contribution of kinetic inductance from aluminum. \textbf{c,} Comparison of quality factors across different types of $\lambda$/4 resonators discussed in this work.}
    \label{fig:meas_Al_shunted}
\end{figure}

We observe the resonant frequency shift in the \ch{NbSe2}-terminated resonator at higher microwave power (Fig.~\ref{fig:measurement}, main text), while the Al-terminated resonator does not show this behavior (Fig. \ref{fig:meas_Al_shunted}b). The resonant frequency shift with increasing photon number in the resonator is described by the Kerr Hamiltonian~\cite{reagor2016superconducting, yurke2006performance, nayfeh2008nonlinear}
\begin{align}
    H = \hbar \omega_\mathrm{r} a\textsuperscript{\textdagger} a + \frac{\hbar}{2} K a\textsuperscript{\textdagger} a\textsuperscript{\textdagger} a a
    \label{eq:Kerr}
\end{align}
where $\hbar$ is the reduced Planck constant, $a$ and $a\textsuperscript{\textdagger}$ are the creation and the annihilation operators, $\omega_\mathrm{r}$ is the resonant frequency, and $K$ is the Kerr constant, which is a measure of the strength of the nonlinearity in the system. The Kerr nonlinearity shifts the resonance frequency and is proportional to the resonator photon number. In addition to the power-dependent frequency shift, we observe a broadening of the internal linewidth as the number of photons in the cavity increases. The Kerr coefficient of \ch{NbSe2} is determined by analyzing the relationship between the resonance frequency shift ($\Delta f$) and the resonator photon number ($n_\mathrm{r}$) in the resonator. We calculate the photon number in the resonator as a function of the applied power using the following formula~\cite{norris2024improved},
\begin{align}
    n_\mathrm{r} = \frac{2 \kappa P_\mathrm{rf}}{\hbar \omega_\mathrm{r} (\kappa + \gamma)^2}
    \label{eq:int_photon}
\end{align}
Here, $\kappa$ represents the coupling rate to the feedline, $P_\mathrm{rf}$ is the power applied at the resonator input, and $\gamma$ is the internal loss rate.

Expanding the current-dependent kinetic inductance relation of Eq.~\ref{eq:Lk_current} (main text), the quartic energy contribution is expressed as
\noindent
\begin{align}
U_\mathrm{m}= \frac{1}{2}L_\mathrm{k} \frac{I_\mathrm{DC}^4}{I^{\ast2}}
\label{eq:Lk_current_energy}
\end{align}
Here, $I^\ast$ sets the scale for the non-linearity of the film. Applying circuit quantization, the current operator can be expanded in terms of ladder operators as $\hat{I_\mathrm{DC}} = I_{\mathrm{zpf}} \left( a + a^\ast \right)$ where $I_{\mathrm{zpf}} = \sqrt{ \frac{ \hbar \omega_\mathrm{r} }{ 2 L_\mathrm{0} }}$ is the zero-point current fluctuation of the mode and $L_\mathrm{0}$ is the total inductance of the resonator mode. We obtain the quantized nonlinear energy term of Eq.~\ref{eq:Lk_current_energy} as
\noindent
\begin{align}
\hat{U}m = \frac{1}{2}L_k \frac{I{\mathrm{zpf}}^4}{I^{\ast 2}} (a + a^\ast)^4
\label{eq:Lk_current_energy2}
\end{align}
The Kerr nonlinearity arises from the term proportional to $a^\ast a^\ast a a$ in the quartic expansion, and the resulting Kerr coefficient ($K$) is
%a\textsuperscript{\textdagger} a\textsuperscript{\textdagger}
\noindent
\begin{align}
%     K = \frac{6L_0 I_{zpf}^4}{\hbar I\ast^2} = \frac{3 \hbar \omega_\mathrm{r}^2 }{2 L_0 I\ast^2}
%     \label{eq:Lk_current_energy3}
% \end{align}
K = \frac{6 L_\mathrm{k} I_{\mathrm{zpf}}^4}{\hbar I^{\ast 2}} = \frac{3 \hbar \omega_\mathrm{r}^2 }{2 L_\mathrm{0}^2} \cdot \frac{L_\mathrm{k}}{I^{\ast 2}}
\label{eq:Lk_current_energy3}
\end{align}

For device D6 (shown in Fig.~\ref{fig:measurement} of the main text), we calculate the Kerr coefficient to be $K/2\pi = -3.8$ Hz/photon using Eq.~\ref{eq:Lk_current_energy3}, which is in good agreement with the value ( $K/2\pi = -1.7$) extracted from the power-dependent frequency shift. The calculation uses $f_\mathrm{r,Al-NbSe_2}$ = \SI{4.79}{GHz}, $L_\mathrm{0} = L_\mathrm{g} + L_\mathrm{k}$ = \SI{2.627}{nH}, and $I^\ast$ = \SI{0.4}{mA}.

\section{Extraction of Kinetic Inductance from Resonance Frequency}
To extract the kinetic inductance ($L_\mathrm{k}$) of \ch{NbSe2}, we employ a comparative model using $\lambda/4$-resonators terminated with aluminum and \ch{NbSe2}. For the aluminum-terminated resonator with a length of $l$ and a resonant frequency of $f_\mathrm{r,Al}$, the input impedance $Z_\mathrm{Al}$ is expressed as \cite{pozar2011microwave}
\begin{equation}
    Z_\mathrm{Al,in} = Z_0 \tanh{(\alpha+i\beta) l} 
    \label{eq:Zin_Al}
\end{equation}
where, $Z_0$ is the characteristic impedance which is 50 $\Omega$ in our work, $\alpha$ and $\beta$ are the attenuation constant and phase constant respectively. For a lossless line, the attenuation constant $\alpha$ is zero, and $\beta (f)$ becomes 
\begin{equation}
    \beta (f) = \frac{2 \pi f_{r,Al} \sqrt{\epsilon_{eff}}}{c}
\end{equation}
 where $\epsilon_{eff}$ is the effective permittivity of the substrate (Si), and $c$ is the speed of light in vacuum. Assuming a low-loss transmission line, we take $\alpha = 0$, so that the input impedance simplifies to $Z_\mathrm{Al,in} = i Z_0 \tan{(\beta l) }$. On resonance, $\beta l = \pi /2$, which yields the resonant frequency of the Al-terminated resonator $f_{r,Al} = \frac{c}{4 l \sqrt{\epsilon_{eff}}}$.
 
For the $\lambda/4$-resonator terminated by \ch{NbSe2}, the kinetic inductance ($L_\mathrm{k}$) contributes an additional frequency-dependent term in the impedance, which is modeled as,
 \begin{equation}
    Z_\mathrm{Al-NbSe_2,in} = Z_0 \frac{i\omega_\mathrm{r,Al-NbSe_2} L_\mathrm{k} + Z_0 \tanh{\left(i\beta l\right)}}{Z_0 + i \omega_\mathrm{r,Al-NbSe_2} L_\mathrm{k} \tanh{\left(i\beta l\right)}}
    \label{eq:Zin_nbse2}
\end{equation}
The physical length $l$ of the resonator is determined from the aluminum reference resonator using Eq.~\ref{eq:Zin_Al}, assuming identical geometry and effective permittivity. On resonance, the impedance of a $\lambda/4$-resonator diverges; for the \ch{NbSe2}-terminated case, Eq.~\ref{eq:Zin_nbse2} reduces to
\begin{equation}
\tan\left( \frac{\omega_\mathrm{r,Al-NbSe_2} \sqrt{\epsilon_{eff}}}{c} l\right) = \frac{Z_0}{\omega_\mathrm{r,Al-NbSe_2} L_\mathrm{k}}
\label{eq:Zin_nbse2_Lk}
\end{equation}
We extract the kinetic inductance $L_\mathrm{k}$ of \ch{NbSe2} from Eq.~\ref{eq:Zin_nbse2_Lk}, where the resonant frequency of \ch{NbSe2}-terminated resonator is in the linear regime (in the single-photon limit).

\section{Kinetic inductance: Clean and Dirty Limits}
% In this section, we consider the theory of kinetic inductance in both the clean and dirty limits for superconductors. 

% The kinetic inductance of a superconductor originates from the inertia of Cooper pairs to the applied field. For a superconducting wire of length $l$ and cross-sectional area $A$, the kinetic energy of the Cooper pairs is $E_\mathrm{k} = \frac{1}{2} \left(\frac{m}{n_\mathrm{s}e^2}\frac{l}{A}\right) I^2$. The term in the parentheses is the general expression of the kinetic inductance.
% \begin{equation}
%     L_\mathrm{k} = \frac{m}{n_\mathrm{s}e^2} \frac{l}{A}
%     \label{eq:Lk_general_1}
% \end{equation}
 
The purity of a superconductor is characterized by the ratio of the mean free path ($l_\mathrm{MFP}$) of the Cooper pairs to the BCS coherence length  ($\xi_0$). The mean free path ($l_\mathrm{MFP}$) is defined as $l_\mathrm{MFP} = \tau v_\mathrm{F}$ with $\tau$ representing the time interval between collisions in the normal state of the superconductor and $v_F$ denoting the Fermi velocity \cite{tinkham2004introduction}. The superconductor is in the dirty metal limit if the mean free path ($l_\mathrm{MFP}$) is much shorter than the BCS coherence length ($\xi_0$), and in the clean limit if $l_\mathrm{MFP}$ is much longer than $\xi_0$ \cite{parks1970superconductivity}. To obtain kinetic inductance for both the dirty and clean limit, two pieces of information are required: (1) the generalized first London equation for a superconductor relating current density and electric field, and (2) an expression characterizing the penetration depth (specifically in the dirty/clean limit). 
The first London equation establishes a direct relation between the supercurrent density ($\Vec{J}$) and the electric field ($\Vec{E}$),  
\begin{equation}
    \frac{\partial \Vec{J}}{\partial t} = \frac{1}{\mu_0 \lambda^2} \Vec{E}
    \label{eq:London}
\end{equation}
where $\lambda$ is the penetration depth and $\mu_0$ is the permeability of free space. When combined with Maxwell’s equations, the London formalism further predicts that magnetic fields decay exponentially inside a superconductor, with the characteristic length $\lambda$. With a sinusoidal drive of frequency $\omega$, Eq.~\ref{eq:London} gives
\begin{subequations}
\begin{align}
    i \omega \left(\mu_0 \lambda^2\right) J = E \label{eq:London_sinusoidal} \\
    i \omega \left(\mu_0 \lambda^2 \frac{l}{A}\right) I = V \label{eq:London_sinusoidal1}
\end{align}
\end{subequations}

% $m$ is the mass of the charge carriers, $n_s$ is the density of superconducting carriers, and $e$ is the electron charge. When combined with Maxwell’s equations, the London formalism further predicts that magnetic fields decay exponentially inside a superconductor, with the characteristic length scale of this screening defined as the London penetration depth,  
% \begin{equation}
%     \lambda = \sqrt{\frac{m}{\mu_0 n_s e^2}}
% \end{equation}
% Here, $\mu_0$ is the permeability of free space. Using the definition of the London penetration depth and assuming a sinusoidal drive, Eq.~\ref{eq:London} is expressed in terms of $\lambda$ as  

% The first London equation establishes a direct correlation between the current density ($\Vec{J}$) and the vector potential ($\textbf{A}$). This directly relates the supercurrent to changes in the magnetic vector potential ($\partial A/\partial t$) and indicates that electric fields ($\Vec{E}$) inside a superconductor are proportional to the rate of change of magnetic fields. \begin{equation}
%     \frac{\partial \left(\Vec{J}\right)}{\partial t} = \frac{n_s e^2}{m} \Vec{E}
%     \label{eq:London}
% \end{equation}
% Here $m$ is the mass of the charge carriers, $n_s$ is the density of the superconducting carriers, and $e$ is the charge of an electron. 

Equation~\ref{eq:London_sinusoidal1} follows from the substitutions $V = El$ and $I = JA$, and the prefactor in the parentheses represents the kinetic inductance. 
\begin{equation}
    L_\mathrm{k} = \mu_0 \lambda^2 \frac{l}{A}
    \label{eq:Lk_general}
\end{equation}
Equation~\ref{eq:Lk_general} depends on the phenomenological parameter $\lambda$, which needs to be evaluated in both clean and dirty limits.

\subsection{Clean Limit}
In the clean limit, where the mean free path of the superconducting carriers is much longer than the coherence length, scattering events can be neglected. In this case, the dynamics of the supercurrent are determined solely by the acceleration of Cooper pairs under the applied electric field. This leads to the relation
\begin{equation}
    \dfrac{d}{dt} (m_\mathrm{eff}\Vec{v}) = \dfrac{d}{dt} \left(\frac{m_\mathrm{eff}}{n_\mathrm{0}e^2}\vec{J}\right)= \vec{E}
    \label{eq:clean}
\end{equation}
Here, $n_\mathrm{0}$ is the carrier density and $m_\mathrm{eff}$ is the effective mass of the supercurrent carriers. Equation~\ref{eq:clean} allows us to identify the penetration depth ($\lambda_\mathrm{clean}$) and kinetic inductance ($L_\mathrm{k,clean}$) in the clean limit as,
\begin{equation}
    \mu_0 \lambda_{clean}^2 = \frac{m_\mathrm{eff}}{n_\mathrm{0} e^2}
\end{equation}
\begin{equation}
    L_\mathrm{k,clean} = \frac{m_\mathrm{eff}}{n_\mathrm{0} e^2} \left(\frac{l}{A}\right)
    \label{eq:Lk_clean}
\end{equation}
This penetration depth is known as the London penetration depth ($\lambda_\mathrm{clean} = \lambda_\mathrm{L}$).

\subsection{Dirty Limit}
In the dirty limit, impurity scattering shortens the mean free path of the carriers, which weakens superfluid screening and causes the magnetic field to penetrate deeper, leading to an enhanced penetration depth given by~\cite{orlando1991,tinkham2004introduction},
\begin{equation}
    \lambda_\mathrm{dirty} \approx \lambda_\mathrm{L} \left( \frac{\xi_0}{l_\mathrm{MFP}}\right)^{1/2}
\end{equation}
where $\xi_0$ is the BCS coherence length at zero temperature $\left(\xi_0 = \frac{\hbar v_\mathrm{F}}{\pi \Delta_0} = \xi_\mathrm{clean}\right)$ and the coherence length in the dirty limit becomes,
\begin{equation}
    \xi_\mathrm{dirty} \approx \xi_0 \left( \frac{l_\mathrm{MFP}}{\xi_0}\right)^{1/2} =  \left(\xi_0 l_\mathrm{MFP} \right)^{1/2}
\end{equation}
Plugging the expression of $\lambda_\mathrm{dirty}$ into the general expression for $L_\mathrm{k}$ in Eq.~\ref{eq:Lk_general}, we get
\begin{equation}
    L_\mathrm{k,dirty} = \mu_0 \lambda_\mathrm{dirty}^2 \frac{l}{A} = \mu_0 \lambda_\mathrm{L}^2 \left(\frac{\xi_0}{l_\mathrm{MFP}}\right) \frac{l}{A}
\end{equation}
Now, using the expression of the BCS coherence length ($\xi_0 = \hbar v_\mathrm{F}/\pi \Delta_0$) and London penetration depth ($\lambda_\mathrm{L}$), we obtain
\begin{equation}
    L_\mathrm{k,dirty} = \frac{m_\mathrm{eff}}{n_\mathrm{0}e^2} \frac{\hbar v_\mathrm{F}}{\pi \Delta_0} \frac{l}{l_\mathrm{MFP} A}
    \label{eq:Lk_dirty}
\end{equation}
Equation~\ref{eq:Lk_dirty} can be further simplified by substituting the expressions for mean free path ($l_\mathrm{MFP} = v_\mathrm{F} \tau $) and resistivity ($\rho_\mathrm{s} = 1/\sigma$).
\begin{equation}
    L_\mathrm{k,dirty} = \rho_\mathrm{s} \frac{\hbar}{\pi \Delta_0} \frac{l}{A} = R_\mathrm{s} \frac{\hbar}{\pi \Delta_0} \frac{l}{w}
    \label{eq:LK_dirty_1}
\end{equation}
where $R_\mathrm{s}$ is the normal-state sheet resistance of a current-carrying superconducting wire of length $l$, width $w$, and film thickness $d$. The BCS superconducting gap is approximated to $\Delta_0 \approx 1.764k_\mathrm{B} T_\mathrm{c}$ for $T \ll T_\mathrm{c}$ with $T_\mathrm{c}$ the critical temperature and Eq.~\ref{eq:LK_dirty_1} becomes
\begin{equation}
    L_\mathrm{k,sq,dirty} = \frac{\hbar}{1.764 k_\mathrm{B}} \frac{R_\mathrm{s}}{T_\mathrm{c}}
    \label{eq:LK_dirty_final}
\end{equation}

% For BCS superconductors, in the low-frequency limit ($hf \ll$ $k_B$T), the Mattis-Bardeen formula for the complex conductivity can be written in terms of the ratio of the imaginary conductivity $\sigma_2$ (recall that the conductivity of the superconducting state can be written in a complex form  $\sigma$ =  $\sigma_1$ - $i\sigma_2$ \cite{glover1956transmission,glover1957conductivity}) to the normal state conductivity $\sigma_\mathrm{n}$,
% \begin{equation}
%         \frac{\sigma_2}{\sigma_\mathrm{n}} = \frac{\pi \Delta}{hf} \tanh \left(\frac{\Delta}{2k_\mathrm{B}T}\right)
% \end{equation}

% In a superconductor, $\sigma_1$ is much less than $\sigma_2$, hence $(\sigma_1 + i\sigma_2)^{-1}$ approximately becomes $(\sigma_1$/$\sigma_2^2$ - $i\sigma_2)$, and $Z_L(\omega)$ = $i\frac{l}{wd}\sigma_2(\omega)^{-1}$ = $i\omega L_k(\omega)$. Therefore, the relation between kinetic inductance, sheet resistance, and the superconducting gap becomes, 
% \begin{equation}
%         L_\mathrm{k,dirty} = \frac{R_\mathrm{s}h}{2\pi^2 \Delta} \frac{1}{\tanh\left(\frac{\Delta}{2k_\mathrm{B}T}\right)} \left(\frac{l}{w}\right) 
%        \label{eq:LK_tan}
% \end{equation}

% We use Eq.~\ref{eq:LK_tan} to estimate the kinetic inductances of \ch{NbSe2} films used in Fig.~\ref{fig:Lk_layer}b.
% \subsection{Kinetic inductance of superconductors in the clean limit}
% \subsection{Comparing the kinetic inductance of \ch{NbSe2} to the clean- and dirty-limit models}

\begin{table}[h]
\centering
\singlespacing
\begin{tabular}{||c | c | c | c | c | c||}\hline
Layers & $L_\mathrm{k,sq}$ & T$_\mathrm{c}$ &  R$_{s}$ & R$_\mathrm{s}$/T$_\mathrm{c}$ & ${L_{k,sq}}/{L_{k,sq,clean}}$\\
 &  (pH/sq)  & (K) &  ($\Omega$/sq) & ($\Omega$/K) & \\
\hline\hline
1 & 1195 & 3.5\textsuperscript{~\cite{khestanova2018unusual}} & 240\textsuperscript{~\cite{khestanova2018unusual}} & 68.57 & 14.95 \\\hline
4 & 282 & 4.8 & 60 & 12.5 & 3.53\\\hline
6 &  191 & 5.5 & 29 & 5.27 & 2.39\\\hline
7 &  143 & 6.0 & 25 & 4.17 & 1.79\\\hline
8 &  135 & 7 & 19 & 2.71 & 1.79\\\hline
8 &  135 & 6.3\textsuperscript{~\cite{cao2015quality}} & 14\textsuperscript{~\cite{cao2015quality}} & 2.22 & 1.75\\\hline
10 &  56 & 7.0\textsuperscript{~\cite{khestanova2018unusual}} & 10\textsuperscript{~\cite{khestanova2018unusual}} & 1.43 & 0.71\\\hline
\end{tabular}
\caption{\textbf{Kinetic inductance of \ch{NbSe_2} from clean to dirty limit}}
\label{table:SI3}
\end{table}

\subsection{Full range calculation: from clean to dirty limit}
\begin{figure}[htbp]
    \centering
    \includegraphics[width=0.8\textwidth]{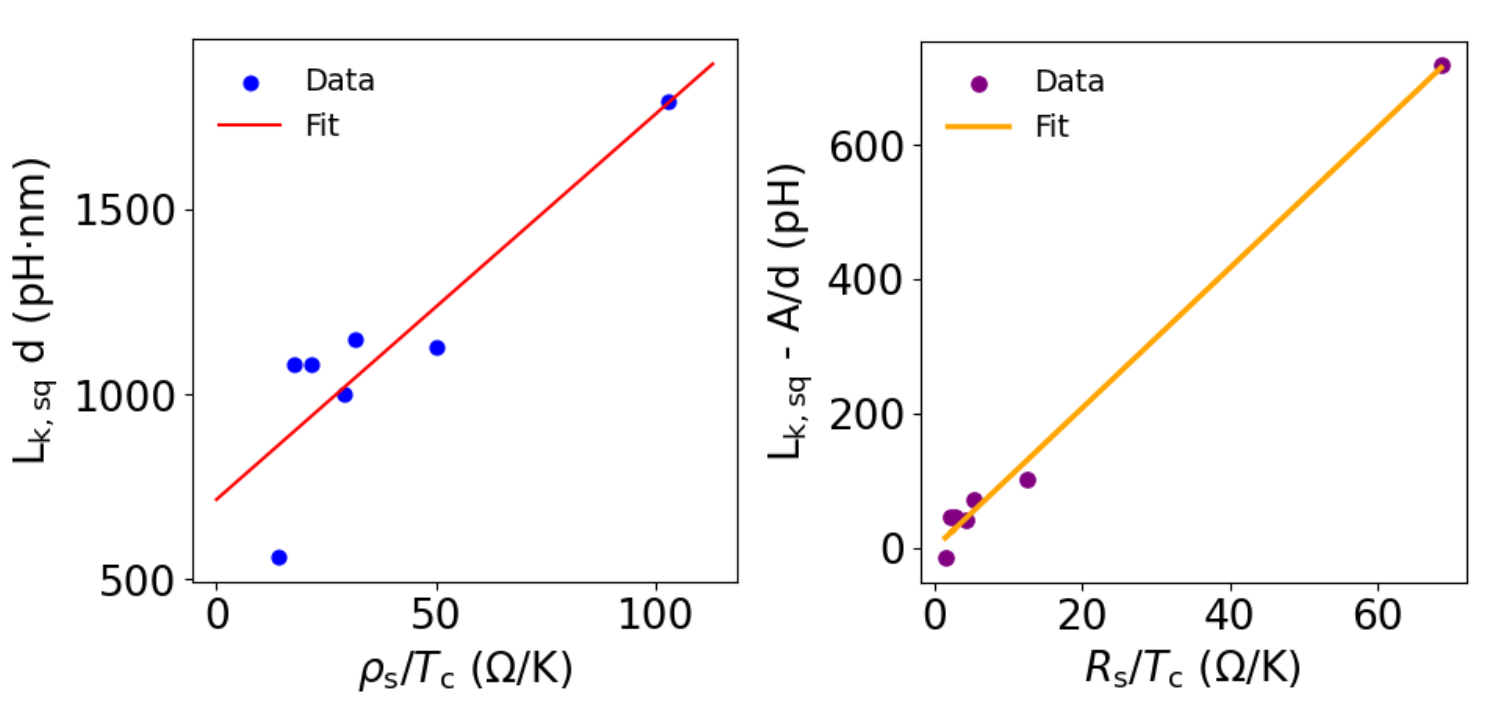}
    \caption{\textbf{a,} The linear relationship of $L_\mathrm{k,sq}\times d$ with $\rho_\mathrm{s}/T_\mathrm{c}$ where the intercept of the y-axis sets the value of the fitting parameter $A = \frac{m_{eff}}{n_0 e^2}$. \textbf{b,} $L_\mathrm{k,sq} - \frac{A}{d}$ is plotted as a function of  $R_\mathrm{s}/T_\mathrm{c}$, where the linear fit shows the dirty limit contribution for the thinner samples of \ch{NbSe_2}.}
    \label{fig:Lk_clean_dirty_analysis_NbSe2}
\end{figure}

We can calculate the kinetic inductance for arbitrary $\xi_0/l_{MFP}$ (where the ratio of $\xi_0/l_{MFP}$ is of the order of 1). The Ginzburg-Landau theory gives a penetration depth of the form which can be well approximated by(Eq. 3.123b in ~\cite{tinkham2004introduction} and the appendix in~\cite{orlando1979critical}): 
\begin{subequations}
\begin{align}
    \lambda = \lambda_\mathrm{L} \left(1+ a \frac{\xi_0}{l_{MFP}} \right)^{1/2} \label{eq:Lambda_total}\\
    L_\mathrm{k} = \mu_0 \lambda_\mathrm{L}^2 \left(1+ a \frac{\xi_0}{l_{MFP}} \right) \left( \frac{l}{A}\right)\label{eq:LK_total}\\
    L_{k,sq} = \mu_0 \frac{\lambda_\mathrm{L}^2}{d} + \mu_0 \frac{\lambda_\mathrm{L}^2}{d} \left(a\frac{\xi_0}{l_{MFP}}\right) \label{eq:LK_sq_total}
\end{align}
\end{subequations}
The factor $a$ in Eq.25a-c is the order of unity ($a = 0.75$), but we have set $a$ to be unity in the main text for simplicity. Equation~\ref{eq:LK_sq_total} indicates that the kinetic inductance per square is the sum of the clean and dirty limit results. We can extract the ratio of coherence length to mean free path ($\frac{\xi_0}{l_{MFP}}$) from the above expression.
\begin{equation}
    \frac{\xi_0}{l_{MFP}} = \frac{L_{k,sq}}{L_{k,sq,clean}} -1
    \label{eq:xi_l}
\end{equation}
Figure~\ref{fig:Lk_clean_dirty_analysis_NbSe2} shows the clean- and dirty-limit contributions to the measured sheet kinetic inductance $L_\mathrm{k,sq}$ of \ch{NbSe_2} samples, highlighting the $1/d$ dependence of the clean-limit term in thicker samples and the dominant dirty-limit contributions in thinner samples.

Figure~\ref{fig:Lk_clean_dirty_analysis_NbSe2} plots results from the second expression in Eq.~\ref{eq:LK_clean_dirty_maintext} which is of the form $L_\mathrm{k,sq} = A/d + (\hbar \ 1.764k_\mathrm{B}) R_\mathrm{s}/T_\mathrm{c}$. Figure~\ref{fig:Lk_clean_dirty_analysis_NbSe2}a plots $L_\mathrm{k,sq} \times d$ versus $\rho/T_\mathrm{c}$ where $\rho = R_\mathrm{s}d$. The y-intercept of Fig.~\ref{fig:Lk_clean_dirty_analysis_NbSe2}a is used to get an estimate of $A$. Figure~\ref{fig:Lk_clean_dirty_analysis_NbSe2}b uses this value of $A$ to plot, $L_\mathrm{k,sq} -A/d$ versus $R_\mathrm{s}/T_\mathrm{c}$, where the y axis now shows the contribution to $L_\mathrm{k,sq}$ from the dirty limit. Given that $A/d$ is about \SI{70}{pH} for the 10 layer sample, the values for  the thicker samples have smaller contributions from the dirty limit term and are thus mostly in the clean limit. The samples for high sheet resistance are dominated by the dirty limit contribution.

\section{Pearl Length}
In superconducting films where the thickness $d$ is much smaller than the penetration depth $\lambda_L$ ($d \ll \lambda$), magnetic fields are weakly screened in the transverse direction and instead spread laterally over long distances. This lateral field penetration is characterized by the Pearl length, defined as
\begin{equation}
        \Lambda = \frac{2 \lambda^2}{d}
       \label{eq:pearl}
\end{equation}

The Pearl lengths of all the devices in this work are estimated using Eq.~\ref{eq:pearl}, where the penetration depth of \ch{NbSe_2} is \SI{230}{nm}.

\section{Determination of \ch{NbSe2} Thickness and Layer Number}
\subsection{AFM characterization of hBN-\ch{NbSe2}-hBN stack}
\begin{figure}[htbp]
    \centering
    \includegraphics[width=0.8\textwidth]{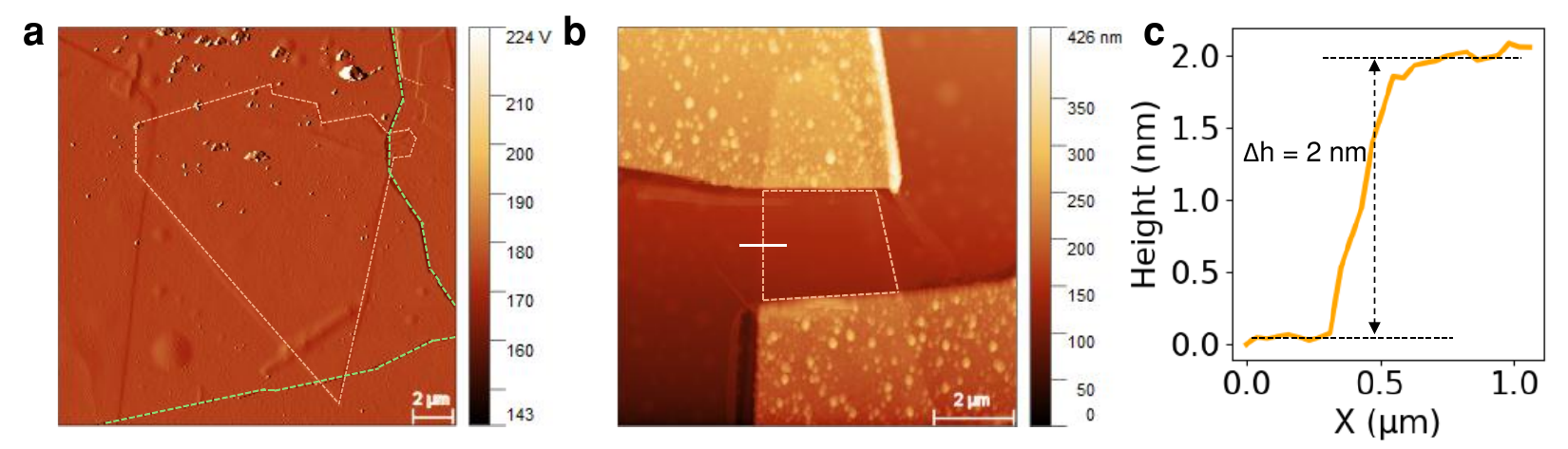}
    \caption{\textbf{AFM topographic image of hBN encapsulated \ch{NbSe2} devices.} \textbf{a,} Amplitude topography of the \ch{NbSe2} flake following the assembly of the hBN-\ch{NbSe2}-hBN heterostructures (hBN-green and \ch{NbSe2}-orange). \textbf{b,} Height profile along the indicated path (white line) to determine the thickness of \ch{NbSe2}.}
    \label{fig:afm}
\end{figure}

Atomic force microscopy (AFM) is a high-resolution imaging technique used in this work to characterize the surface topography and height of hBN-\ch{NbSe2}-hBN stacks. The thickness of hBN encapsulated \ch{NbSe2} is determined by tapping-mode AFM, illustrated in Fig.~\ref{fig:afm}. The AFM measurement tends to overestimate the actual thickness of \ch{NbSe2} due to the presence of the top hBN encapsulation. In this work, the effective \ch{NbSe2} thickness is slightly smaller, as confirmed by the shear mode in the low-frequency Raman spectrum. 

\subsection{Raman Spectroscopy of thin \ch{NbSe2} films }
\begin{figure}[htbp]
    \centering
    \includegraphics[width=0.8\textwidth]{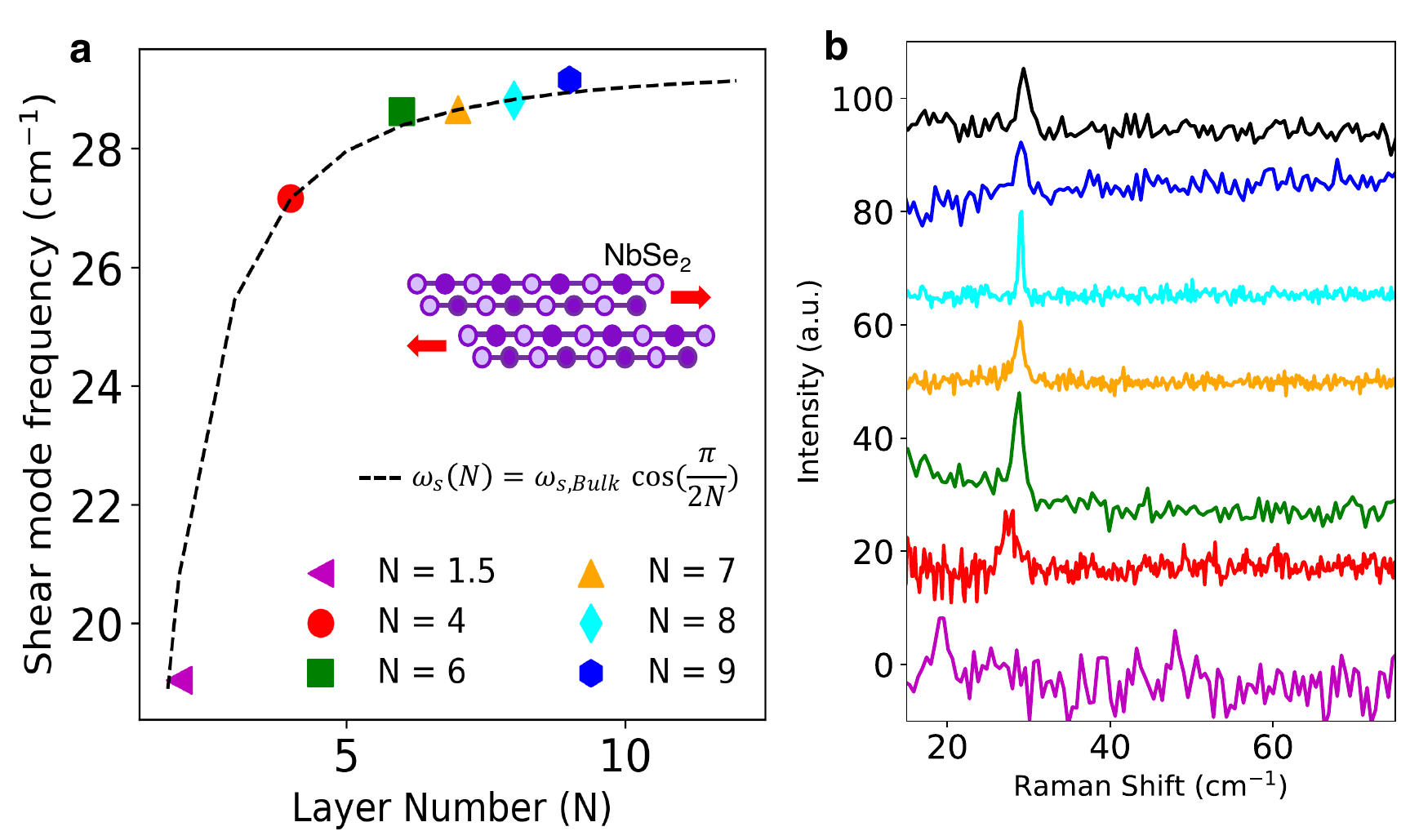}
    \caption{\textbf{Low-frequency Raman spectroscopy on different layers of \ch{NbSe2} samples at room temperature under vacuum for the perpendicular polarization configuration.} \textbf{a,} Plot of shear mode freqeuency versus layer numbers of \ch{NbSe2}. The dotted blue line shows the $\omega_s = \omega_{s,Bulk} \cos{(\frac{\pi}{2N})}$ relation with $\omega_{s,Bulk}$ = \SI{29.39}{cm\textsuperscript{-1}}. \textbf{b,} Low-frequency Raman spectra of \ch{NbSe2} with layer number N = 4 – 8 and bulk \ch{NbSe2}. For clarity, the spectra are vertically displaced. }
    \label{fig:Raman_all}
\end{figure}

The number of layers in our samples was determined by the shear mode frequency of Raman spectroscopy ~\cite{xi2015strongly, he2016interlayer}. The Raman data were collected using a Horiba LabRAM Evolution system in the XY channel at room temperature. A 50x objective (Mitutoyo MY50X-825) was employed to direct and collect light in the backscattering geometry with a 532 nm wavelength. Data were captured using a liquid nitrogen-cooled CCD, maintained at -133$^\circ$C. The incident laser power on the sample was estimated to be below 4 mW. Figure~\ref{fig:Raman_all}b shows that the low-frequency mode observed in the Raman spectra corresponds to interlayer shear modes. In all spectra, shear modes are analyzed using Lorentzian peak fitting after subtracting the background. The shear mode peak in Raman spectra originates from interlayer shearing and is absent in the monolayer sample.
The frequency of the shear modes ($\omega_\mathrm{s}$) decreases with decreasing layer number ($N$), reflecting a reduced effective interlayer spring constant. The relationship can be described as $\omega_\mathrm{s} = \omega_\mathrm{s,Bulk} \cos{(\frac{\pi}{2N})}$ with a bulk shear mode frequency $\omega_\mathrm{s,Bulk}$ = \SI{29.39}{cm\textsuperscript{-1}} \cite{he2016interlayer, lui2014temperature}. 
The AFM-measured thicknesses of the \ch{NbSe2} flakes are consistent with the layer numbers determined from the shear mode frequency (1 layer = \SI{1}{\nm}) for all samples except for the thinnest device D1 (see main text table). For D1, the AFM measurement indicates a thickness of \SI{2}{nm}, while Raman spectroscopy only shows a weak feature at the frequency \SI{19.04}{cm\textsuperscript{-1}} rather than a well-defined Lorentzian peak (Fig~\ref{fig:Raman_all}b). This corresponds to a layer number less than 2, with a fitted value of N =1.5 (Fig.~\ref{fig:Raman_all}a). We use this Raman-derived value (\SI{1.5}{\nm}) in the analysis presented in Fig.~\ref{fig:Lk_layer}a.

\pagebreak

\begin{figure}[htbp]
    \centering
    \includegraphics[width=0.8\textwidth]{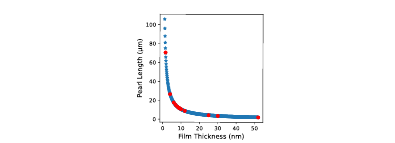}
    \caption{\textbf{Pearl length estimation (blue) of \ch{NbSe2} of different thicknessess with consideration of $\lambda_L$ = 230 nm.} The \ch{NbSe2} samples of this work are identified in red circles. }
    \label{fig:Pearl_NbSe2}
\end{figure}

\end{document}